\documentclass{raa_twocolumn} 

\usepackage{graphicx,times}             
\usepackage{natbib}
\usepackage{amssymb,amsmath}
\bibpunct{(}{)}{;}{a}{}{,}
\usepackage[pagebackref=true]{hyperref}
\usepackage{xcolor}

\begin{document}

  \title{First Gamma-Ray Burst Observations with \textit{SVOM}}

   \volnopage{Vol.0 (202x) No.0, 000--000}      
   \setcounter{page}{1}                         

   \author{
      Frédéric Daigne\inst{1,2}
   \and
   Damien Turpin\inst{3}
    \and
   Jean-Luc Atteia\inst{4}
   \and
    Jesse T. Palmerio\inst{3}
   \and
   Li-Ping Xin\inst{5,6}
   \and
   Susanna D. Vergani\inst{7,1}
   \and
    Bing Zhang\inst{8,9} 
   \and
   Chao Wu\inst{5,6} 
   \and
   Tais Maiolino\inst{10} 
   \and
   Benjamin Schneider\inst{11} 
   \and
   Donghua Zhao\inst{5}
   \and
   An Li\inst{12,13} 
   \and
   Daniele Malesani\inst{14,15,16} 
   \and
   Hua Li Li\inst{5}
   \and
   Andrea Saccardi\inst{3,17}
   \and
   Maria-Grazia Bernardini\inst{18,10} 
   \and
   He Gao\inst{12,13} 
   \and 
   Frédéric Piron\inst{10} 
   \and
   Olivier Godet\inst{4}
   \and
   Bin-Bin Zhang\inst{19} 
   \and
   Yun Wang\inst{20} 
   \and
    Liangjun Chen\inst{21} 
   \and
   Bertrand Cordier\inst{3}
   \and
   Jian-Yan Wei\inst{5,6}
   \and
   Stéphane Basa\inst{11} 
   \and
   Arnaud Claret\inst{3}
   \and
   Alexis Coleiro\inst{22} 
   \and
   Jin-Song Deng\inst{5,6}
   \and
   Yong-Wei Dong\inst{23} 
   \and
   Diego Götz\inst{3}
   \and
   Xu-Hui Han\inst{5,6}
   \and
   Yu-Lei Qiu\inst{5}
   \and
   Cyril Lachaud\inst{22} 
   \and
   En-Wei Liang\inst{21} 
   \and
   Jing Wang\inst{5,6}
   \and
   Shao-Lin Xiong\inst{23} 
   \and
   Shuang-Nan Zhang\inst{23} 
   on behalf of the \textit{SVOM} collaboration\inst{24}.
   }

   \institute{
        Sorbonne Université, CNRS, UMR 7095, Institut d'Astrophysique de Paris, 98 Bis bd Arago, 75014, Paris, France; {\it daigne@iap.fr}; \\
        \and
        Institut Universitaire de France; 
        \and
        CEA Paris-Saclay, IRFU/DAp-AIM, 91191 Gif sur Yvette, France;
        \and
        IRAP, Université de Toulouse, CNRS, CNES, Toulouse, France;
        \and
        National Astronomical Observatories, Chinese Academy of Sciences, Beijing 100101, China;
        \and
        School of Astronomy and Space Science, University of Chinese Academy of Sciences, Beijing 101408, China;
        \and
        LUX, Observatoire de Paris, Université PSL, CNRS, Sorbonne Université, Meudon, 92190, France;
        \and
        The Hong Kong Institute for Astronomy and Astrophysics, the University of Hong Kong, Pokfulam Road, Hong Kong, P. R. China;
        \and
        Department of Physics,  the University of Hong Kong, Pokfulam Road, Hong Kong, P. R. China;
        \and
        Laboratoire Univers et Particules de Montpellier, Université Montpellier, CNRS/IN2P3, F-34095 Montpellier, France;
        \and
        Aix Marseille University, CNRS, CNES, LAM, Marseille, France;
        \and
        Institute for Frontier in Astronomy and Astrophysics, Beijing Normal University, Beijing 102206, China;
        \and
        Department of Astronomy, Beijing Normal University, Beijing 100875, China;
        \and
        Niels Bohr Institute, University of Copenhagen, Jagtvej 155, 2200, Copenhagen N, Denmark;
        \and 
        The Cosmic Dawn Centre (DAWN), Denmark;
        \and
        Department of Astrophysics/IMAPP, Radboud University, PO Box 9010, 6500 GL, The Netherlands;
        \and
        Centre national d’études spatiales (CNES), Paris, France;
        \and
        INAF – Osservatorio Astronomico di Brera, Via Bianchi 46, I23807 Merate, LC, Italy;
        \and 
        School of Astronomy and Space Science, Nanjing University, Nanjing 210093, China;        
        \and 
        Purple Mountain Observatory, Chinese Academy of Sciences, Nanjing 210023, China;
        \and
        Guangxi Key Laboratory for Relativistic Astrophysics, School of Physical Science and Technology, Guangxi University, Nanning 530004, China;  
        \and 
        Université Paris Cité, CNRS, Astroparticule et Cosmologie, F75013 Paris, France;       
        \and
        Key Laboratory of Particle Astrophysics, Institute of High Energy Physics, Chinese Academy of Sciences, Beijing 100049, China;
        \and
        \url{https://fsc.svom.org/home/collaboration/collaborators}.\\
\vs\no
  {\small Received 2026 January 21; accepted 2026 June 21}}

\abstract{
Following its launch on 22 June 2024, the Space-based multi-band astronomical Variable Objects Monitor (\textit{SVOM}) successfully completed its flight acceptance, commissioning, and scientific validation phases in early 2025, during which several tens of gamma-ray bursts (GRBs) were detected onboard.
Three quarters of these events have also been detected by other satellites, and a quarter are \textit{SVOM}-only GRBs. In this article, we describe these early GRB observations, with a first description of the \textit{SVOM} GRB sample that is emerging, and of the level of characterisation already achieved, and with a focus on a few events of particular interest. 
These early results are very encouraging regarding \textit{SVOM}'s ability to detect and fully characterise 
(including prompt emission, afterglow and distance) a wide range of GRBs (classical long GRBs, short GRBs, X-Ray Flashes, etc.) and to enable the use of these extreme high-energy transients as probes of the distant Universe.
}

   \authorrunning{F. Daigne, D. Turpin~et~al.}              
   \titlerunning{First GRB Observations with \textit{SVOM}} 

   \maketitle


\section{Introduction}          
\label{sect:intro}

The core program of \textit{SVOM} (Space-based multi-band astronomical Variable Objects Monitor) is devoted to Gamma-Ray Burst (GRB) studies \citep{svom_white_paper}. 
The characteristics of the mission (space and ground-based instruments, slewing capability, observation strategy following an alert, etc.) were mainly shaped by this scientific objective \citep{Cordier+etal+2026a}. The prime objective
is to build a sample of fully characterised GRBs, including events belonging to the various classes of GRBs: classical long GRBs (LGRB), short GRBs (SGRB) and other GRBs associated to neutron star mergers, soft GRBs classified as X-ray Flashes (XRF) or X-Ray Rich GRBs (XRR), etc. The goal 
is in particular to characterise for the same events the temporal and spectral properties of the prompt emission and multi-wavelength afterglow, and to measure the redshift (distance). When possible, the aim is also to characterise the host galaxy and to search for other electromagnetic  (supernovae, kilonovae) or multi-messenger (gravitational waves, neutrinos) counterparts. 
Such a sample will help us advance our understanding of GRB progenitors, of the nature of their central engines, of the properties of their relativistic ejecta, and of the dissipation mechanisms and radiative processes responsible for the different emission phases.
This requires in particular a fast and efficient follow-up including spectroscopy at large telescopes, and when possible late observations of the host galaxies. These are also the key observations required to use GRBs as probes of the distant Universe, the second main scientific goal of \textit{SVOM} core program. 

This article presents early results of this program, for the period from launch 
to March 31, 2025 corresponding mainly to the commissioning and scientific validation phase. 
The first  observations by \textit{SVOM} of other astrophysical sources  not related to GRBs are reported in \citet{Coleiro+etal+2026}.
During this period, the various components of \textit{SVOM} were gradually put in place, and by October 2024, the complete sequence for observing a GRB (detection, localization, slew, follow-up) with rapid communication to the community via the NASA General Coordinates Network (GCN,  \citealt{barthelmy:1998}) was in place. However, the configuration continued to improve gradually, with, in particular, a lowering of the slew threshold in December 2024, the implementation of automatic Target of Opportunity (ToO) requests for the X-ray follow-up of GRBs detected and localized by the coded mask telescope ECLAIRs on board \textit{SVOM} with  the X-Ray Telescope (XRT, \citealt{burrows2005}) on board \textit{Swift} in February 2025, and the Follow-up X-ray Telescope (FXT) on board \textit{Einstein Probe} (\textit{EP}; \citealt{yuan2022}) in April 2025.  The GRB detection rate by ECLAIRs and the Gamma-Ray Monitor (GRM) on board SVOM during this period is discussed in Section~\ref{sect:triggers}.
The already excellent efficiency of the afterglow detection and follow-up and redshift determination for GRBs localized by ECLAIRs is presented in Section~\ref{sect:followup}. Section~\ref{sect:diversity} shows with a few examples that these early detections   
are very encouraging regarding
the capacity of \textit{SVOM} to fully characterise GRBs of various types 
and to develop the use of GRBs as cosmological probes, and Section~\ref{sect:conclusion} is the conclusion.

   
\section{GRB Detection Rates On Board \textit{SVOM}}
\label{sect:triggers}

Two wide-field of view instruments can trigger on GRBs on board \textit{SVOM}: the coded mask telescope ECLAIRs working in photon counting mode in the 4-150 keV range, with a field of view of 2 sr 
\citep{Godet+etal+2026a}, and the Gamma-Ray Monitor GRM in the 15 keV-5 MeV range, with a field of view of 2.6 sr for each of its three Gamma-Ray detectors (GRDs, \citealt{Sun+etal+2026}).
During the 9.3 first months of the \textit{SVOM} mission, from the launch on 22 June 2024 to the end of the scientific validation phase at the end of March 2025, 
86 GRBs were detected on board \textit{SVOM}\footnote{There were also 18 on-ground detections, not discussed here.}, including 36 GRBs also detected by 
the Gamma-ray Burst Monitor (GBM, \citealt{meegan2009}) on board \textit{Fermi}, 16 by 
the Burst Alert Telescope (BAT, \citealt{barthelmy2005}) on board 
\textit{Swift},
10 by Konus \citep{aptekar1995}
on board \textit{Wind},
 and 7 by the Wide field X-ray Telescope (WXT) on board \textit{EP} \citep{yuan2022}. About one quarter
  of these early GRBs were detected only by \textit{SVOM}.

As illustrated in several examples in Section~\ref{sect:diversity}, the combination of ECLAIRs and GRM allows to measure the spectrum of the GRB prompt emission from 4 keV to 5 MeV. In
the future, it is expected that the large field of view of the 
\textit{SVOM} Ground-based Wide Angle Camera (GWAC)
will allow
to extend the characterization of the prompt emission to the optical range in a few cases 
\citep{Xin+etal+2026}. We now focus on the GRB detections by the two gamma-ray instruments on board the \textit{SVOM} satellite.

\begin{figure}
   \centering
   \includegraphics*[width=0.85\linewidth, angle=0]{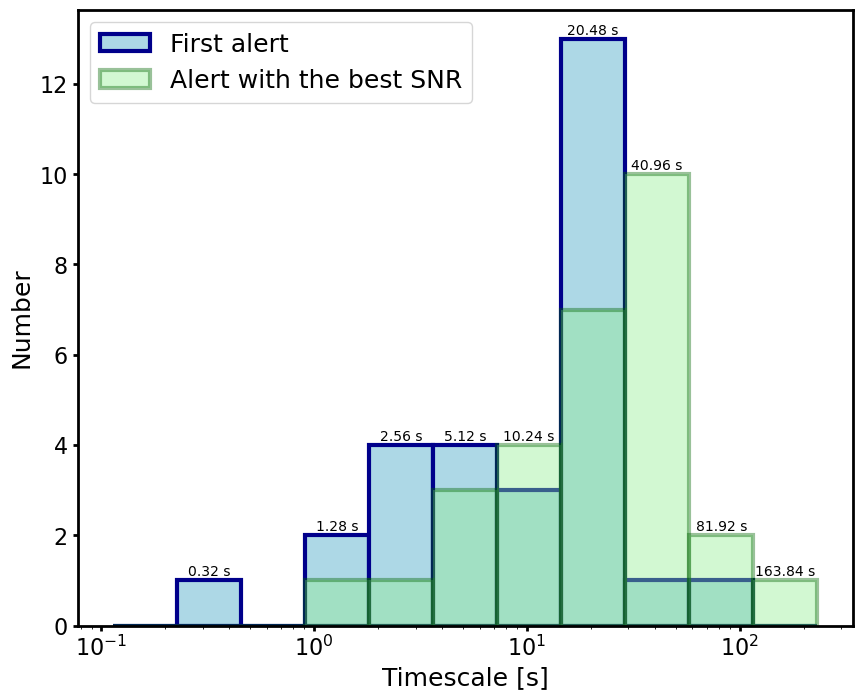}
   \includegraphics[width=0.85\linewidth, angle=0]{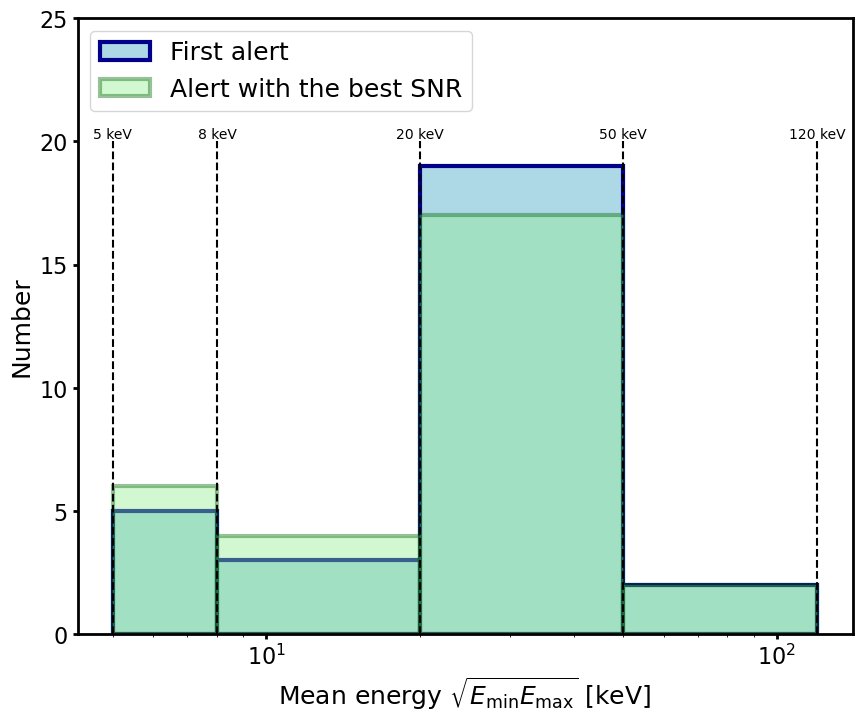}
   \caption{\textbf{ECLAIRs onboard trigger.} Distribution of the timescale (top) and energy band (bottom) of the first and best SNR alert for the first 29 GRBs detected onboard by ECLAIRs.}
   \label{fig:stat_trig_ecl}
\end{figure}

\subsection{ECLAIRs Onboard Triggers}
ECLAIRs can trigger on many combinations of timescales, energy bands and zones in the detector plane, using either
the count rate (Count Rate Trigger or CRT, searching for a count excess over background, on time intervals from 10 ms to 20 s) or 
images (Image Trigger or IMT, searching for a transient sources in sky images on timescales from 20 s to 20 min)
\citep{Schanne+etal+2026}. 
For most GRBs, several
alert packets are thus produced on board \textit{SVOM} and then transmitted immediately to the
ground thanks to the \textit{SVOM} very high frequency (VHF) network
\citep{Cordier+etal+2026b}.
The CRT trigger is always followed by 
the reconstruction of its sky image to search for a transient source, yielding
the reported signal-to-noise
(SNR) in the initial notice and circular
that are sent to the NASA GCN to report the detection of a new GRB.
During the first 9.3 months following the launch, 29 GRBs were detected by ECLAIRs on board \textit{SVOM}, including 18 \textit{SVOM}-only triggers. All these GRBs were localized within a few arcmin,  
with a median statistical error of 7', 
to which a 2' systematic error must be added in quadrature.
Figure~\ref{fig:stat_trig_ecl}  shows for these 29 first GRBs the distribution of the timescale and energy band of the  
alert packet received first by the VHF network (hereafter "first alert")
and of the 
alert packet with the best SNR among all those received (hereafter "best SNR alert"). The impact of the longest timescales (to be compared to the maximum timescale of 32 s  used by \textit{Swift}/BAT, \citealt{barthelmy2005}), is clearly seen, with 45\% (resp. 7\%) of the best SNR alerts (resp. first alerts) corresponding to large
timescales (40.96, 81.92 and 163.84 s). 
The low-energy threshold at 4 keV of ECLAIRs is significantly lower than that of previous missions, which has also a clear impact: in 21\% of cases, the best SNR alert is obtained in the 5-8 keV
energy band. As discussed below, this gives access to the poorly known population of softer GRBs (XRRs or XRFs). As the fraction of time with an active onboard trigger has increased from 44\% up until November 2024 to 75\% from December 2024 to March 2025, these early detections lead to an expected rate of at least 48-52 GRBs per year detected and localized by ECLAIRs on board \textit{SVOM} during the scientific operations.  

\subsection{GRM Onboard Triggers}

The GRM can trigger on three timescales (0.1, 1 and 4~s), only if the signal is above threshold in at least two of the three GRDs and only if ECLAIRS has not triggered first. This led to 77 GRBs detected onboard in the first 9.3 months, among which $26\%$ are also detected by ECLAIRs. This corresponds to an expected rate of $\sim 100$ GRBs per year detected onboard, and an expected total rate of $\sim 130$ GRBs per year taking into on-ground triggers and ECLAIRs first triggers. 
GRBs detected by both instruments ECLAIRs and GRM 
offer the best characterisation of the prompt emission, with a spectral coverage from 4~keV to 5 MeV.

\defcitealias{GCN_z_240821A}{37731}
\defcitealias{GCN_Gemini_240821A}{37319}
\defcitealias{GCN_XRT_240821A}{37249}
\defcitealias{GCN_VT_240821A}{37243}
\defcitealias{GCN_FXT_240821A}{37316}
\defcitealias{GCN_GRM_240821A}{37226}
\defcitealias{GCN_ECL_240821A}{37220}
\defcitealias{GCN_GBM_240821A}{37219}

\defcitealias{GCN_FXT_241001A}{37809}
\defcitealias{GCN_XRT_241001A}{37725}
\defcitealias{GCN_VT_241001A}{37695}
\defcitealias{GCN_UVOT_241001A}{37678}
\defcitealias{GCN_z_241001A}{37677}
\defcitealias{GCN_LCO_241001A}{37673}
\defcitealias{GCN_ECL_241001A}{37655}
\defcitealias{GCN_JWST_241001A}{37867}

\defcitealias{GCN_z_241209B}{38646}
\defcitealias{GCN_FXT_241209B}{38632}
\defcitealias{GCN_Konus_241209B}{38537}
\defcitealias{GCN_XRT_241209B}{38525}
\defcitealias{GCN_VT_241209B}{38516}
\defcitealias{GCN_ECLGRM_241209B}{38478}

\defcitealias{GCN_z_250103B}{38820}
\defcitealias{GCN_VT_250103B}{38813}
\defcitealias{GCN_ECLGRM_250103B}{38797}
\defcitealias{GCN_BATXRT_250103B}{38796}

\defcitealias{GCN_z_241217A}{38637}
\defcitealias{GCN_WXTFXT_241217A}{38606}
\defcitealias{GCN_VT_241217A}{38600}
\defcitealias{GCN_XRT_241217A}{38599}
\defcitealias{GCN_ECLGRMMXT_241217A}{38594}

\defcitealias{GCN_UVOT_241030B}{38008}
\defcitealias{GCN_z_241030B}{38004}
\defcitealias{GCN_GRM_241030B}{38002}
\defcitealias{GCN_VT_241030B}{37999}
\defcitealias{GCN_Nanshan_241030B}{37985}
\defcitealias{GCN_ECL_241030B}{37984}
\defcitealias{GCN_BATXRT_241030B}{37981}
\defcitealias{GCN_GBM_241030B}{37980}

\defcitealias{GCN_XRT_250329A}{39924}
\defcitealias{GCN_z_250329A}{39918}
\defcitealias{GCN_VT_250329A}{39917}
\defcitealias{GCN_ECLGRMMXT_250329A}{39916}
\defcitealias{GCN_Colibri_250329A}{39915}

\defcitealias{GCN_UVOT_250327B}{40009}
\defcitealias{GCN_Konus_250327B}{39978}
\defcitealias{GCN_Colibri_250327B}{39897}
\defcitealias{GCN_XRT_250327B}{38895}
\defcitealias{GCN_z_250327B}{38893}
\defcitealias{GCN_VT_250327B}{38890}
\defcitealias{GCN_ECLMXT_250327B}{38888}

\defcitealias{GCN_z_250317B}{39769}
\defcitealias{GCN_XRT_250317B}{39763}
\defcitealias{GCN_Colibri_250317B}{39754}
\defcitealias{GCN_ECLMXTVTUL_250317B}{39752}

\defcitealias{GCN_NOT_250317B}{39770}
\defcitealias{GCN_OHP_250317B}{39767}
\defcitealias{GCN_REM_250317B}{39764}

\defcitealias{GCN_GBM_250205A}{39171}
\defcitealias{GCN_WXTFXT_250205A}{39165}
\defcitealias{GCN_Colibri_250205A}{39162}
\defcitealias{GCN_XRT_250205A}{39161}
\defcitealias{GCN_z_250205A}{39160}
\defcitealias{GCN_LT_250205A}{39156}
\defcitealias{GCN_ECLMXT_250205A}{39154}

\defcitealias{GCN_XRT_250103A}{38849}
\defcitealias{GCN_zNOT_250103A}{38814}
\defcitealias{GCN_zGTC_250103A}{38809}
\defcitealias{GCN_GBM_250103A}{38806}
\defcitealias{GCN_VT_250103A}{38802}
\defcitealias{GCN_CGFTUL_250103A}{38791}
\defcitealias{GCN_ECLGRMMXTUL_250103A}{38786}

\defcitealias{GCN_FXT_250314A}{39739}
\defcitealias{GCN_XRT_250314A}{39734}
\defcitealias{GCN_z_250314A}{39732}
\defcitealias{GCN_MXTUL_250314A}{39729}
\defcitealias{GCN_VTUL_250314A}{39728}
\defcitealias{GCN_NOT_250314A}{39727}
\defcitealias{GCN_ECLGRM_250314A}{39719}

\defcitealias{GCN_ECL_250328A}{39910}
\defcitealias{GCN_MXT_250328A}{39967}
\defcitealias{GCN_VT_250328A}{39936}
\defcitealias{GCN_Colibri_250328A}{39920}
\defcitealias{GCN_CGFT_250328A}{39920}
\defcitealias{GCN_z_250328A}{44399}

\defcitealias{svom_paper_250317B_inprep}{Zhao~et~al. (in prep.)}
\defcitealias{svom_paper_250327B_inprep}{Yang~et~al. (in prep.)}
\defcitealias{svom_paper_250205A_inprep}{Saccardi~et~al. (in prep.)}
\defcitealias{svom_paper_240821A_inprep}{Daigne~et~al. (in prep.)}
\defcitealias{svom_paper_241217A_inprep}{Brunet~et~al. (in prep.)}
\defcitealias{svom_paper_250129A_inprep}{Chen~et~al. (in prep.)}

\begin{table}
    \begin{center}
    \caption[]{\label{tab:followup_efficiency} \textbf{Efficiency of the Follow-Up of the First \textit{SVOM} GRBs.} The table reports the number of afterglow detections in X-rays and visible/near infrared (NIR) (by \textit{SVOM} instruments or others, see the details for the first 13 ECLAIRs GRBs with a measured redshift in Table~\ref{tab:bestsample}), as well as the number of measured redshifts for the  GRBs detected on board \textit{SVOM} during the 9.3 months following the launch: GRBs with a GRM-only onboard trigger and GRBs with an ECLAIRs onboard trigger, with a focus on the cases where the trigger was followed by an automatic slew of the satellite.}
    \resizebox{\linewidth}{!}{\begin{minipage}{1.05\linewidth}
    \begin{tabular}{lcccc}
        \hline\noalign{\smallskip}
        & \# & \multicolumn{2}{c}{Afterglow} & Redshift\\
        & & X-rays & Visible/NIR & \\
        \noalign{\smallskip}\hline
        \hline\noalign{\smallskip}
        GRM-only & 57 & 17 (30\%) & 12 (21\%) & 11 (19\%)\\
        \noalign{\smallskip}\hline\noalign{\smallskip}
        ECLAIRs (all) & 29 & 28 (97\%) & 20 (69\%) & 13 (45\%) \\
        ECLAIRs (slew) & 16 & 16 (100\%) & 13 (81\%) & 9 (56\%)\\
        \noalign{\smallskip}\hline
    \end{tabular}
    \end{minipage}}
    \end{center}
\end{table}

\begin{figure*}
   \centering
   \includegraphics[width=0.9\linewidth]
   {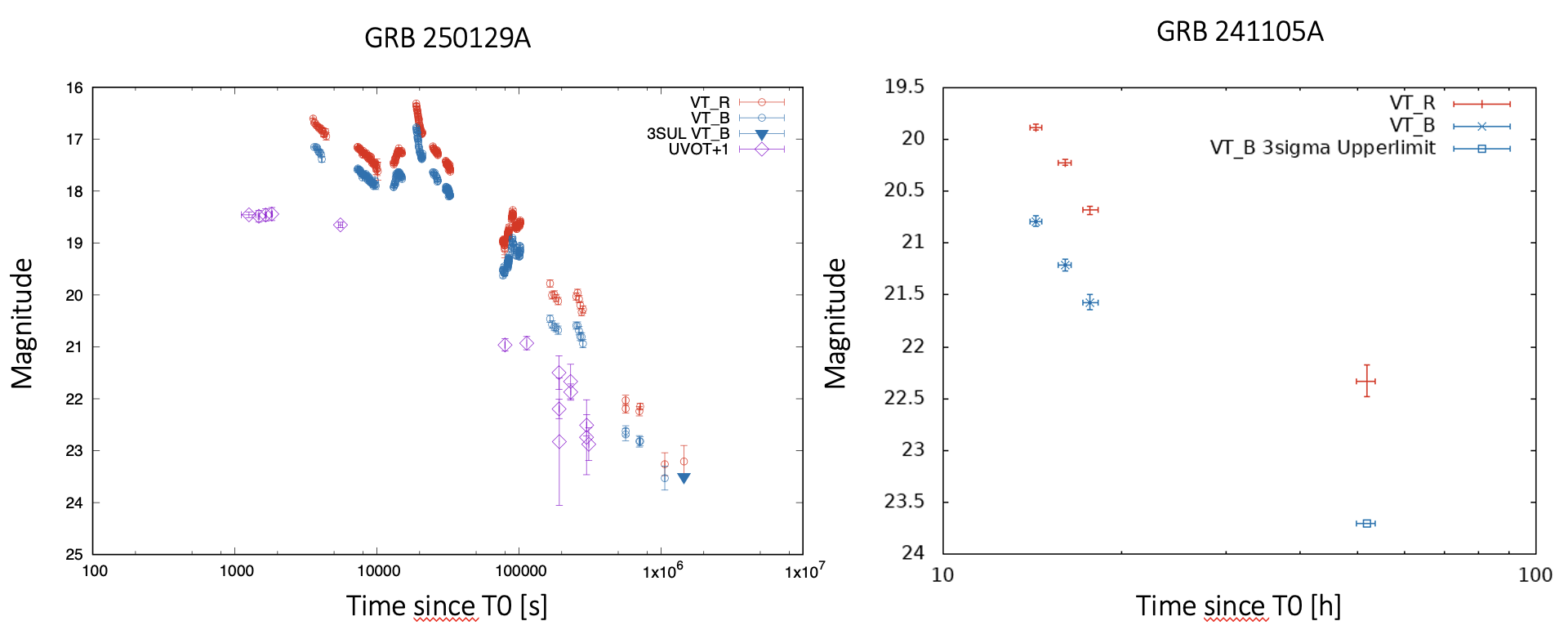}
   \caption{\textbf{Follow-up observations of two \textit{Swift}/BAT GRBs by \textit{SVOM}/VT.}
    Left: VT observations of the afterglow of the long GRB250129A (red and blue symbols). \textit{Swift}/UVOT observations (purple) are shown for comparison. Figure adapted from \citetalias{svom_paper_250129A_inprep}.
    Right: VT observations of the afterglow of GRB241105A, contributing to the multi-wavelength follow-up from X-rays to radio discussed in \citealt{svom_paper_241105a_type3}.}
   \label{fig:vt_250129A_241105A}
\end{figure*}


\section{characterising \textit{SVOM} GRBs: Afterglow, Host Galaxy and Redshift}
\label{sect:followup}

\subsection{An Already Efficient Follow-Up Strategy}

\textit{SVOM} has developed a complex observational strategy to fully characterise the GRBs detected and localized onboard. 
This strategy involves (i)  a fast transmission of the GRB alerts to the ground via a VHF network
\citep{Cordier+etal+2026b},
(ii) the slew of the satellite to detect, localize and follow the afterglow in X-rays with
the Microchannel X-ray Telescope (MXT)
\citep{Goetz+etal+2026}
and in the visible with the
Visible Telescope (VT)
\citep{Qiu+etal+2026}, 
(iii) a rapid reaction with 
\textit{SVOM} on-ground follow-up telescopes
(GFTs), i.e. the C-GFT in China \citep{Wu+etal+2026}, and COLIBRI (also called FM-GFT) in Mexico \citep{Basa+etal+2026},
favored by the anti-solar pointing of \textit{SVOM}, 
and (iv) a fast dissemination of the alerts to the community via notices and circulars on the NASA GCN \citep{Claret+etal+2026,Louvin+etal+2026}.
The follow-up of \textit{SVOM} GRBs benefits 
from
several partnerships with other facilities. 
This includes automatic ToO requests for a follow-up in X-rays by \textit{Swift}/XRT (since February 2025) and \textit{EP}/FXT (since April 2025).
After the immediate reaction to an alert, internal ToOs are used for the long-term follow-up of the afterglow with \textit{SVOM}/VT, allowing in many cases to get a deep and well-sampled lightcurve, as illustrated below by a few examples. \textit{SVOM} ToOs \citep{Cordier+etal+2026a} are also used to react to external GRB alerts. In 2025, this procedure allowed the first detection of the optical counterpart by \textit{SVOM}/VT in 8 cases, where the initial localization was provided by \textit{EP}/WXT or \textit{Swift}/BAT. 

The \textit{SVOM} follow-up strategy was gradually implemented after the launch. Automatic slews started only in October 2024, leading to a fraction of 55\% of GRBs with an onboard ECLAIRs trigger followed by a slew during the first 9.3 months, which has increased to 73\% since December 2024 when the ECLAIRs threshold SNR to request
a slew was lowered. 
Table~\ref{tab:followup_efficiency} summarizes the afterglow detections and redshift measurements that were achieved in the first 9.3 months following GRB detections on board \textit{SVOM}. 

\begin{figure*}
   \centering
   \includegraphics[width=0.95\textwidth]{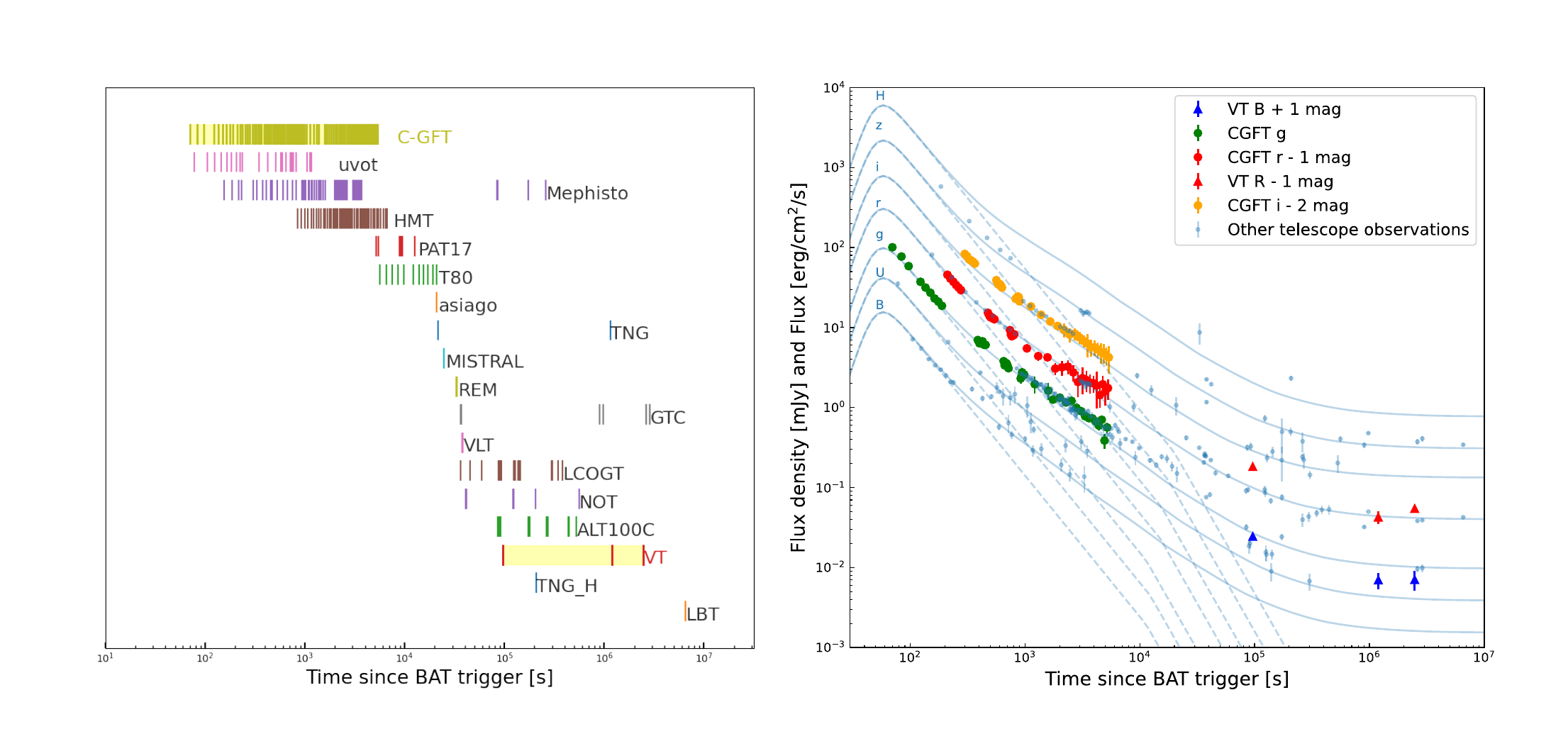}
   \caption{\textbf{Follow-up observations of GRB240825A by \textit{SVOM}/C-GFT and \textit{SVOM}/VT.}
   Left: timeline of the observations of this GRB in the visible (see \citet{svom_paper_240825A} and references therein). \textit{SVOM}/C-GFT provides the earliest observations.
   Right: resulting afterglow lightcurve, adapted from \citet{svom_paper_240825A}, where the two component model (forward and reverse shock) shown in blue lines is discussed.}
   \label{fig:240825A}
\end{figure*}

\subsubsection{\textit{SVOM} GRM-only GRBs Localized by External Missions}
\label{sec:GRM-only_GRBs}
The full characterisation up to the redshift determination is achieved in 19\% of the 57 GRM-only GRBs, corresponding to common triggers with \textit{Swift}/BAT (8/11 GRBs)
and
\textit{EP}/WXT (3/11 GRBs).
The already 
significant impact of \textit{SVOM} on 
the afterglow follow-up of GRBs detected and localized by other satellites, especially thanks to the sensitivity of the VT, is illustrated with two examples in Figure~\ref{fig:vt_250129A_241105A}.
GRB250129A (left 
panel) is a long GRB
at $z=2.151$ \citep{GCN_Stargate_GRB250129A},  detected
by \textit{Swift}/BAT \citep{GCN_BAT_GRB250129A}. 
   The initially bright 
   visible
   counterpart was followed by the
   Ultra-Violet/Optical Telescope (UVOT, \citealt{roming2005}) of \textit{Swift}, 
   covering the early phase
  from  0.3 to 4.7~h after the trigger \citep{GCN_UVOT_GRB250129A}, and a second epoch up to 3~days.
   \textit{SVOM}/VT observations 
   started less than 1~h after the trigger, and lasted for more than two weeks, revealing a complex variable lightcurve, with several flares. Combined with the X-ray lightcurve, which shows also
   flares, 
   these observations will bring
   new constraints on the origin of the
   strong variability in the afterglow phase (\citetalias{svom_paper_250129A_inprep}, see also \citealt{2026arXiv260308555A}).
GRB241105A (right panel)
is a short burst with extended emission, detected
by \textit{Fermi}/GBM \citep{GCN_241105A_Fermi} and \textit{Swift}/BAT \citep{GCN_241105A_BAT-Guano}. \textit{SVOM}/VT observations from 14.2~h to 2.16 days reveal a rapidly decaying lightcurve of the afterglow initially detected by the Gravitational-wave Optical Transient Observer (GOTO)
\citep{GCN_241105A_GOTO} and undetected by \textit{Swift}/UVOT. 
This is a significant contribution to the characterisation of this GRB belonging to a rare class, a fraction of which could be associated to mergers (see  \S~\ref{sect:short}). In this specific case, the high redshift and the properties of the host galaxy rather point towards a collapsar origin  \citep{svom_paper_241105a_type3}.

In addition to \textit{SVOM}/VT, the follow-up of external GRBs also strongly benefits from
the two \textit{SVOM}/GFTs. This is illustrated in Figure~\ref{fig:240825A} for GRB240825A, a classical long GRB detected by \textit{Fermi}/GBM and \textit{Swift}/BAT \citep{GCN_240825A_GBM,GCN_240825A_BAT}. \textit{SVOM}/C-GFT started to observe only 66 s after the trigger. This constitutes the earliest optical observations within a very rich multi-wavelength dataset, enabling
a detailed modelling of the afterglow including the early contribution of the reverse shock \citep{svom_paper_240825A}. 

\begin{table}
    \begin{center}
    \caption[]{\label{tab:bestsample} \textbf{The First  Fully Characterised GRBs Detected by \textit{SVOM}/ECLAIRs.}}
    \vspace*{6ex}

    \resizebox{0.93\linewidth}{!}{\rotatebox{90}{\begin{minipage}{30cm}
    \begin{tabular}{c|c|cccc|cc|cccccc|cccccc}
    \hline\noalign{\smallskip}
    GRB  & Preliminary & \multicolumn{4}{l|}{Prompt: \textit{SVOM}} &  \multicolumn{2}{l|}{Prompt: other}  & \multicolumn{6}{l|}{X-ray afterglow} & \multicolumn{6}{l}{Visible/Near Infrared afterglow}\\ 
    Name & classification & ECLAIRs & GCN & GRM & GCN & Inst. & GCN & MXT & GCN & XRT & GCN & FXT & GCN  & VT & GCN & GFTs & GCN & Other & GCN \\ 
    \noalign{\smallskip}\hline\noalign{\smallskip}
    240821A & SGRB+EE & 
    X  & {\footnotesize\citetalias{GCN_ECL_240821A}} & 
    X  & {\footnotesize\citetalias{GCN_GRM_240821A}} &
    GBM  & {\footnotesize\citetalias{GCN_GBM_240821A}} &
    & & 
    X  & {\footnotesize\citetalias{GCN_XRT_240821A}} &
    X  & {\footnotesize\citetalias{GCN_FXT_240821A}} &
    UL & {\footnotesize\citetalias{GCN_VT_240821A}} &
    & & 
    Gemini N & {\footnotesize\citetalias{GCN_Gemini_240821A}}\\
    241001A & XRF &
    X  & {\footnotesize\citetalias{GCN_ECL_241001A}} &
    & & 
    & & 
    & & 
    X  & {\footnotesize\citetalias{GCN_XRT_241001A}} &
    X  & {\footnotesize\citetalias{GCN_FXT_241001A}} &
    X  & {\footnotesize\citetalias{GCN_VT_241001A}} &
    & & 
    LCO & {\footnotesize\citetalias{GCN_LCO_241001A}}\\
    241209B & LGRB & 
    X  & {\footnotesize\citetalias{GCN_ECLGRM_241209B}} &
    X  & {\footnotesize\citetalias{GCN_ECLGRM_241209B}} &
    Konus & {\footnotesize\citetalias{GCN_Konus_241209B}} &
    & & 
    X  & {\footnotesize\citetalias{GCN_XRT_241209B}} &
    X  & {\footnotesize\citetalias{GCN_FXT_241209B}} &
    X  & {\footnotesize\citetalias{GCN_VT_241209B}} &
    & & & \\
    250103B & LGRB & 
    X  & {\footnotesize\citetalias{GCN_ECLGRM_250103B}} &
    X  & {\footnotesize\citetalias{GCN_ECLGRM_250103B}} &
    BAT  & {\footnotesize\citetalias{GCN_BATXRT_250103B}} &
    & & 
    X  & {\footnotesize\citetalias{GCN_BATXRT_250103B}} &
    & & 
    X  & {\footnotesize\citetalias{GCN_VT_250103B}} &
    & & & \\
    241217A & LGRB & 
    X  & {\footnotesize\citetalias{GCN_ECLGRMMXT_241217A}} &
    X  & {\footnotesize\citetalias{GCN_ECLGRMMXT_241217A}} &
    WXT  & {\footnotesize\citetalias{GCN_WXTFXT_241217A}} &
    X  & {\footnotesize\citetalias{GCN_ECLGRMMXT_241217A}} &
    X  & {\footnotesize\citetalias{GCN_XRT_241217A}} &
    X  & {\footnotesize\citetalias{GCN_WXTFXT_241217A}} &
    X  & {\footnotesize\citetalias{GCN_VT_241217A}} &
    & & & \\
    241030B & LGRB &
    X  & {\footnotesize\citetalias{GCN_ECL_241030B}} &
    X  & {\footnotesize\citetalias{GCN_GRM_241030B}} &
    GBM,BAT  & {\footnotesize\citetalias{GCN_GBM_241030B,GCN_BATXRT_241030B}} &
    & & 
    X & {\footnotesize\citetalias{GCN_BATXRT_241030B}} &
    & & 
    X & {\footnotesize\citetalias{GCN_VT_241030B}} &
    & & 
    Nanshan & {\footnotesize\citetalias{GCN_Nanshan_241030B}} \\
    250329A & LGRB &
    X & {\footnotesize\citetalias{GCN_ECLGRMMXT_250329A}} &
    X & {\footnotesize\citetalias{GCN_ECLGRMMXT_250329A}} &
    & & 
    X & {\footnotesize\citetalias{GCN_ECLGRMMXT_250329A}} &
    X & {\footnotesize\citetalias{GCN_XRT_250329A}} &
    & & 
    X & {\footnotesize\citetalias{GCN_VT_250329A}} &
    COLIBRI & {\footnotesize\citetalias{GCN_Colibri_250329A}} &
    & \\
    250327B & LGRB &
    X & {\footnotesize\citetalias{GCN_ECLMXT_250327B}} &
    X & {\footnotesize\citetalias{svom_paper_250327B_inprep}} &
    Konus & {\footnotesize\citetalias{GCN_Konus_250327B}} &
    X & {\footnotesize\citetalias{GCN_ECLMXT_250327B}} &
    X & {\footnotesize\citetalias{GCN_XRT_250327B}} &
    & & 
    X & {\footnotesize\citetalias{GCN_VT_250327B}} &
    COLIBRI & {\footnotesize\citetalias{GCN_Colibri_250327B}} &
    & \\
    250317B & XRF &
    X & {\footnotesize\citetalias{GCN_ECLMXTVTUL_250317B}} &
    & & 
    & &
    X & {\footnotesize\citetalias{GCN_ECLMXTVTUL_250317B}} &
    X & {\footnotesize\citetalias{GCN_XRT_250317B}} &
    & & 
    X & {\footnotesize\citetalias{svom_paper_250317B_inprep}} &
    COLIBRI & {\footnotesize\citetalias{GCN_Colibri_250317B}} &
    & \\
    250205A & LGRB &
    X & {\footnotesize\citetalias{GCN_ECLMXT_250205A}} &
    X & {\footnotesize\citetalias{svom_paper_250205A_inprep}} &
    GBM,WXT & {\footnotesize\citetalias{GCN_GBM_250205A,GCN_WXTFXT_250205A}} &
    X & {\footnotesize\citetalias{GCN_ECLMXT_250205A}} &
    X & {\footnotesize\citetalias{GCN_XRT_250205A}} &
    X & {\footnotesize\citetalias{GCN_WXTFXT_250205A}} &
    X & {\footnotesize\citetalias{svom_paper_250205A_inprep}} &
    COLIBRI & {\footnotesize\citetalias{GCN_Colibri_250205A}} &
    LT & {\footnotesize\citetalias{GCN_LT_250205A}} \\
    250103A & LGRB &
    X & {\footnotesize\citetalias{GCN_ECLGRMMXTUL_250103A}} &
    X & {\footnotesize\citetalias{GCN_ECLGRMMXTUL_250103A}} &
    GBM & {\footnotesize\citetalias{GCN_GBM_250103A}} &
    UL & {\footnotesize\citetalias{GCN_ECLGRMMXTUL_250103A}} &
    X & {\footnotesize\citetalias{GCN_XRT_250103A}} &
    & & 
    X & {\footnotesize\citetalias{GCN_VT_250103A}} &
    C-GFT (UL) & {\footnotesize\citetalias{GCN_CGFTUL_250103A}} &
    &\\
    250314A & LGRB &
    X & {\footnotesize\citetalias{GCN_ECLGRM_250314A}} &
    X & {\footnotesize\citetalias{GCN_ECLGRM_250314A}} &
    & & 
    UL & {\footnotesize\citetalias{GCN_MXTUL_250314A}} &
    X & {\footnotesize\citetalias{GCN_XRT_250314A}} &
    X & {\footnotesize\citetalias{GCN_FXT_250314A}} &
    UL & {\footnotesize\citetalias{GCN_VTUL_250314A}} &
    & & 
    NOT & {\footnotesize\citetalias{GCN_NOT_250314A}} \\
    250328A & LGRB & 
    X & {\footnotesize\citetalias{GCN_ECL_250328A}}& 
    X & & 
    & & 
    & & 
    X & {\footnotesize\citetalias{GCN_MXT_250328A}}& 
    & & 
    X & {\footnotesize\citetalias{GCN_VT_250328A}}& 
    COLIBRI & 
    {\footnotesize\citetalias{GCN_Colibri_250328A}} &\\
    \noalign{\smallskip}\hline\noalign{\smallskip}
    \end{tabular}
    X: detected; UL:upper limit. Instruments: BAT: \textit{Swift}/BAT ; GBM: \textit{Fermi}/GBM ; Konus: Konus-Wind ; WXT: \textit{EP}/WXT ; XRT: \textit{Swift}/XRT ; FXT: \textit{EP}/FXT.\\
    In the last columns ("Other"), the first detection of the visible/near infrared afterglow is reported when it is not by a \textit{SVOM} telescope: Gemini North (Gemini N), Las Campanas Observatory (LCO), Liverpool Telescope (LT), Nordic Optical Telescope (NOT).\\
    \vspace*{1.5ex}

    \begin{tabular}{c|ccl|cl|cl}
    \hline\noalign{\smallskip}
    GRB & \multicolumn{3}{l|}{Redshift} & \multicolumn{2}{l|}{Host galaxy} & \multicolumn{2}{l}{Supernova} \\
    Name & $z$ & Telescope & GCN & Detection & Reference & Detection & Reference\\
    \noalign{\smallskip}\hline\noalign{\smallskip}
    240821A & $0.238$ & VLT & {\footnotesize \citetalias{GCN_z_240821A}} & GTC & Daigne et al. in prep. & & \\
    241001A & $0.573$ & VLT & {\footnotesize \citetalias{GCN_z_241001A}} & UL (Legacy Survey) & \cite{Dey2019} & JWST & {\footnotesize GCN~\citetalias{GCN_JWST_241001A}} \\
    241209B & $0.575$ & GTC & {\footnotesize \citetalias{GCN_z_241209B}} & UL (Legacy Survey) & \cite{Dey2019} & &\\
    250103B & $1.416$ & VLT & {\footnotesize \citetalias{GCN_z_250103B}} & Legacy Survey & \cite{Dey2019} & &\\
    241217A & $1.879$ & VLT & {\footnotesize \citetalias{GCN_z_241217A}} & Legacy Survey & \cite{Dey2019} & &\\
    241030B & $2.82$ & GTC & {\footnotesize \citetalias{GCN_z_241030B}} & UL (Pan-STARRS1) & \cite{chambersPanSTARRS1Surveys2016} & &\\ 
    250329A & $2.91$ & VLT & {\footnotesize \citetalias{GCN_z_250329A}} & UL (Legacy Survey) & \cite{Dey2019} & & \\
    250327B & $3.035$ & NOT & {\footnotesize \citetalias{GCN_z_250327B}} & UL (Legacy Survey) & \cite{Dey2019} & &\\
    250317B & $3.44$ & GTC & {\footnotesize \citetalias{GCN_z_250317B}} & UL (Legacy Survey) & \cite{Dey2019} & &\\
    250205A & $3.55$ & GTC & {\footnotesize \citetalias{GCN_z_250205A}} & UL (Legacy Survey) & \cite{Dey2019} & & \\
    250103A & $4.01$ & GTC, NOT & {\footnotesize \citetalias{GCN_zGTC_250103A,GCN_zNOT_250103A}} & Legacy Survey & \cite{Dey2019} & & \\ 
    250314A & $\simeq7.3$ & VLT & {\footnotesize \citetalias{GCN_z_250314A}} & JWST & {\footnotesize \citet{jwst_paper_250314A}} & JWST & {\footnotesize \citet{jwst_paper_250314A}} \\
    250328A & $1.250$ & VLT & {\footnotesize \citetalias{GCN_z_250328A}} & VLT & {\footnotesize \citetalias{GCN_z_250328A}} & & \\
    \noalign{\smallskip}\hline
    \end{tabular}\\
    Redshift/Telescope: telescope where the redshift was measured. VLT: Very Large Telescope; GTC: Gran Telescopio Canarias; NOT: Nordic Optical Telescope. 
    \end{minipage}}}
    \end{center}
    \vspace*{-3ex}

\end{table}

\subsubsection{\textit{SVOM} GRBs Detected by ECLAIRs}

As expected, the best follow-up efficiency is achieved for ECLAIRs onboard triggers followed by an automatic slew, which corresponds to 16 GRBs during the first 9.3 months. The X-ray afterglow was detected for all these events, thanks to the MXT (8/16), the \textit{Swift}/XRT (16/16) and \textit{EP}/FXT\footnote{Automatic ToO requests for \textit{EP}/FXT have started right after the beginning of the observation period discussed in the present article.} (6/16). The visible
or near-infrared (NIR) afterglow was detected in 81\% of cases, with 12/16 detections by the VT (and 4 early deep upper limits), and 4/16 detections by the \textit{SVOM}/GFTs (and 3 early upper limits). Finally, the redshift of the GRB was measured in 56\% of cases thanks to \textit{SVOM} partners, 
especially with the Very Large Telescope (VLT) at the European Southern Observatory (7/13; Stargate collaboration), the Gran Telescopio Canarias (GTC; 5/13) and the Nordic Optical Telescope (NOT; 2/13).
Several of these well characterised GRBs are discussed in the next Section.

As only 13 of the 20  ECLAIRs GRBs with a detected visible/NIR afterglow have  a known redshift (see Table~\ref{tab:followup_efficiency}), the redshifts of some of the remaining GRBs may still be measured in the future via late host galaxy spectroscopic observations, as
already done in one case (GRB250328A, \citealt{GCN_z_250328A}).
Among the 4 cases with early deep upper limits by the VT, one GRB has been 
confirmed as a high-redshift GRB (see \S~\ref{sect:cosmo}).

\subsubsection{Optimizing the Follow-Up Strategy}

As discussed in more details in \citet{svom_paper_250314A},
there are a number of well identified key factors where progress can still be made to further improve this already very good follow-up efficiency.
In particular, (i) the compliance with
the nominal pointing law avoiding the Galactic plane has been much better since the start of scientific operations than in the early months; (ii) the fraction of automatic slews following ECLAIRs triggers has increased since December 2024 and can still be optimized; (iii) the automatic ToO requests to \textit{Swift}/XRT and \textit{EP}/FXT following ECLAIRs triggers is now fully operational.

Progress on some other aspects will take longer: 
(i)~the delay to identify possible counterpart in early VT images or to compute  early deep upper limits that may identify high-redshift candidates  could be reduced, especially if the number of available X-band stations is increased to download the full data more rapidly to the ground;
(ii) early NIR photometry, allowing photometric redshift measurements or the detection of high-redshift events, should be enhanced in late 2026 thanks to the new camera CAGIRE (CApturing Gamma-ray bursts Infra-Red Emission) observing in  J and H bands at COLIBRI \citep{Basa+etal+2026}.

\begin{figure}
   \centering
   \includegraphics[width=0.9\linewidth, angle=0]{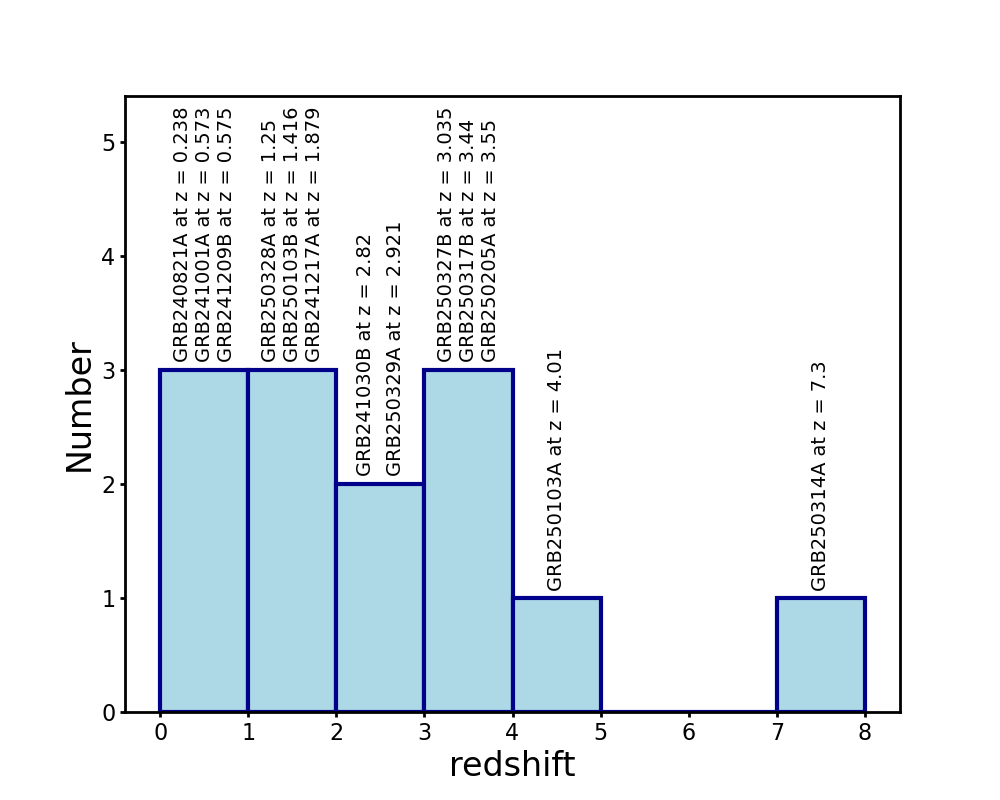}
   \caption{\textbf{Redshift distribution of the first 13 ECLAIRs GRBs with a measured redshift}
    (references are provided in Table~\ref{tab:bestsample}).}
   \label{fig:stat_redshift_ecl}
\end{figure}

\subsection{Redshift and Other Properties of Fully Characterised \textit{SVOM} GRBs}
\label{sec:characterization}

As illustrated in  Section~\ref{sect:diversity}, the prompt emission of the first 29 ECLAIRs GRBs is well characterised. They correspond either to very soft events detected only by ECLAIRs (8/29) or classical GRBs detected by ECLAIRs+GRM, with a spectral coverage from 4 keV to 5 MeV. 
The afterglow is detected in X-rays and visible/NIR for more than two thirds of this sample, and already 
13/29 (45\%) of GRBs have a redshift measurement. These thirteen GRBs and their properties are listed in Table~\ref{tab:bestsample} and their redshift distribution is displayed in Figure~\ref{fig:stat_redshift_ecl}.
This still small sample covers a broad range of redshift,  from $z=0.238$ to $z\simeq 7.3$, and
includes one short GRB with extended emission, two  soft GRBs that can probably be classified as XRFs, and ten long classical GRBs with a median redshift of $2.9$. 
As shown in Table~\ref{tab:bestsample}, their  host galaxy was detected in 6/13 cases. In the 7/13 other cases, an upper limit for the host galaxy was obtained using the galaxy catalogs from the Legacy Survey  
and Pan-STARRS1. Finally, an associated supernova was detected in two cases, thanks to  
follow-up observations with the James Webb Space Telescope (JWST).
As illustrated with a few examples in Section~\ref{sect:diversity}, this 
already shows the capacity of \textit{SVOM} to fully characterise 
a wide diversity of GRBs, constraining their progenitors and  probing their physics, and to use GRBs to study the high-redshift Universe.

\begin{figure}
    \centering
    \includegraphics*[viewport=0cm 0cm 15cm 11cm, width=0.9\linewidth, angle=0]{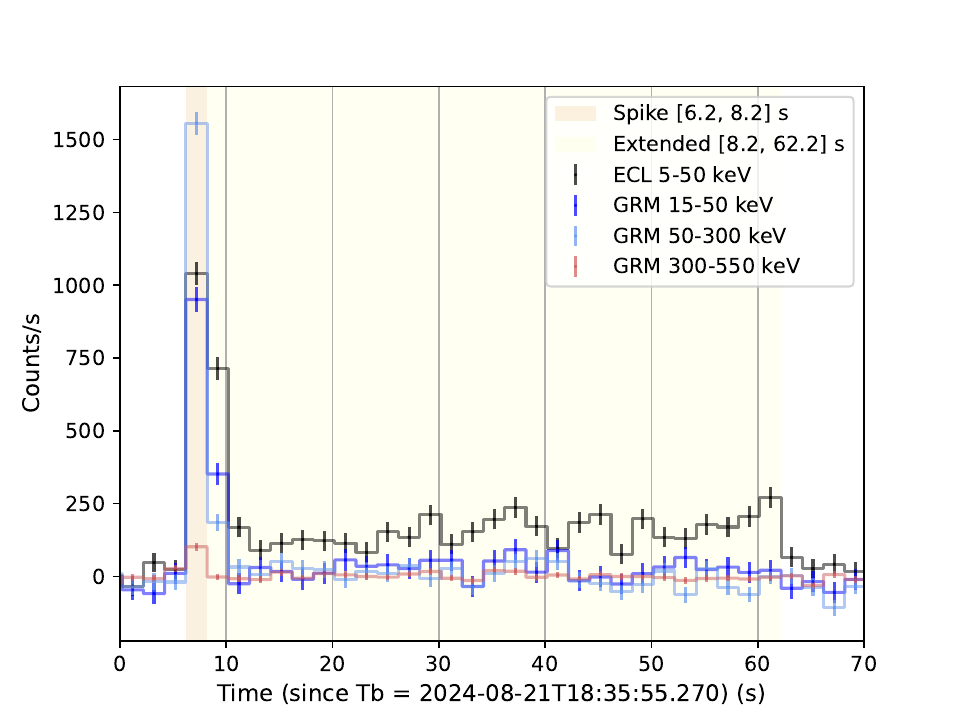}
    \includegraphics*[viewport=0cm 0cm 15cm 11cm,width=0.9\linewidth, angle=0]{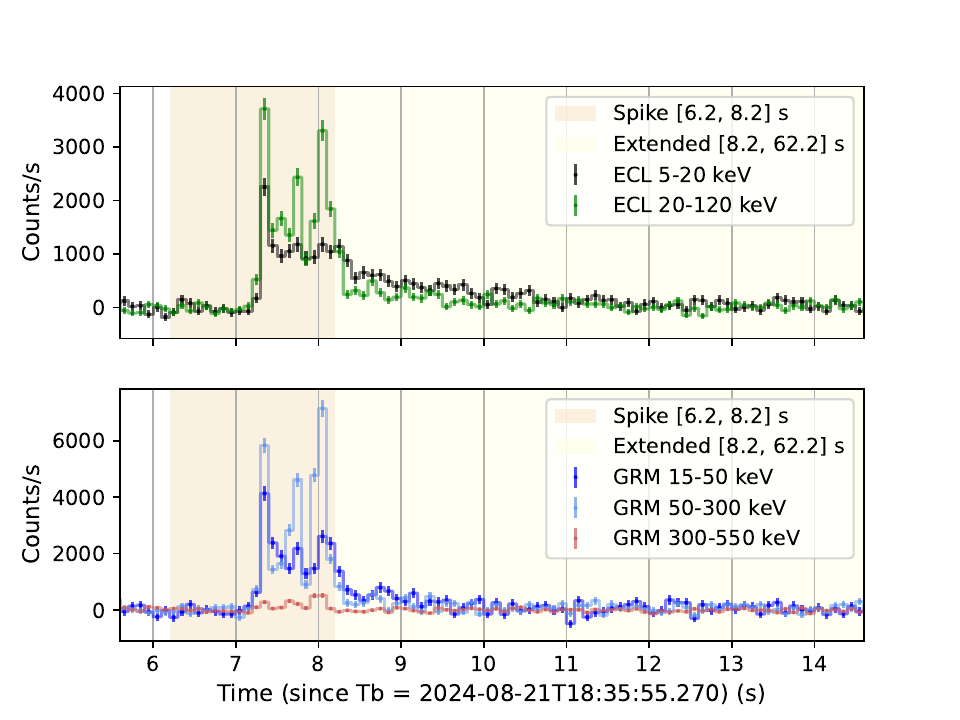}    
    \caption{\textbf{GRB240821A, a fully characterised short GRB with extended emission: lightcurve.}
    Top: ECLAIRs and GRM background-subtracted lightcurves over the whole duration of the GRB, using a time binning of 2 s.
    Middle and bottom: ECLAIRs and GRM  lightcurves, focusing on the first phase of the GRB, with a time binning of 0.1 s. In all panels, the time intervals used for the spectral analysis of the spike and the extended emission in Figure~\ref{fig:ecl_grm_240821A_sp} are indicated.
    Figures adapted from \citetalias{svom_paper_240821A_inprep}.}
   \label{fig:ecl_grm_240821A_lc}
\end{figure}


\section{Exploring the Diversity of the GRB Population with \textit{SVOM}}
\label{sect:diversity}

\begin{figure}
    \centering      
    \centerline{\includegraphics*[viewport=0.5cm 0.9cm 27cm 20.5cm, width=0.85\linewidth, angle=0]{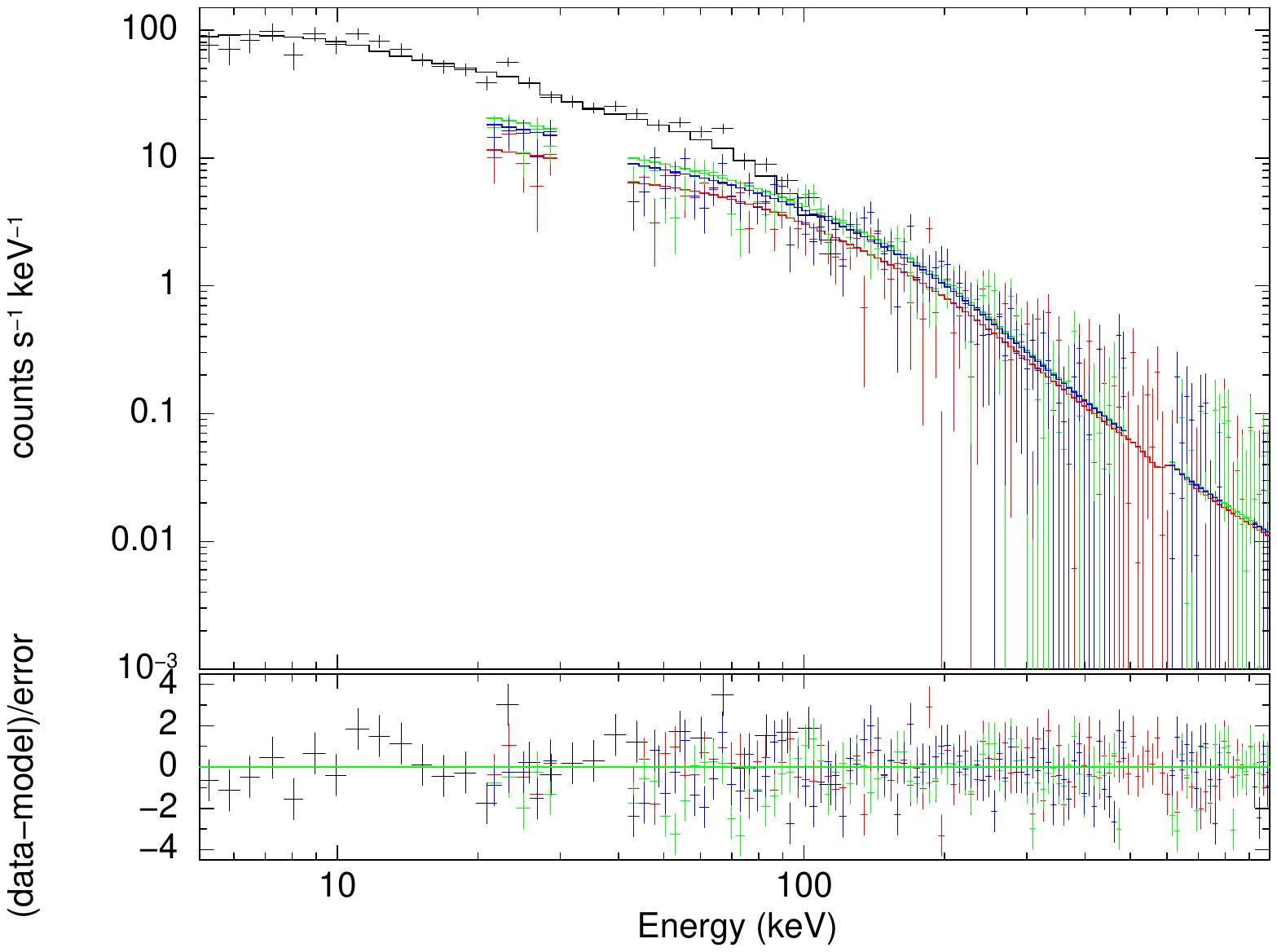}} 
    \vspace*{2ex}
    \centerline{\includegraphics[viewport=0.5cm 0.9cm 27cm 20.5cm, width=0.85\linewidth, angle=0]{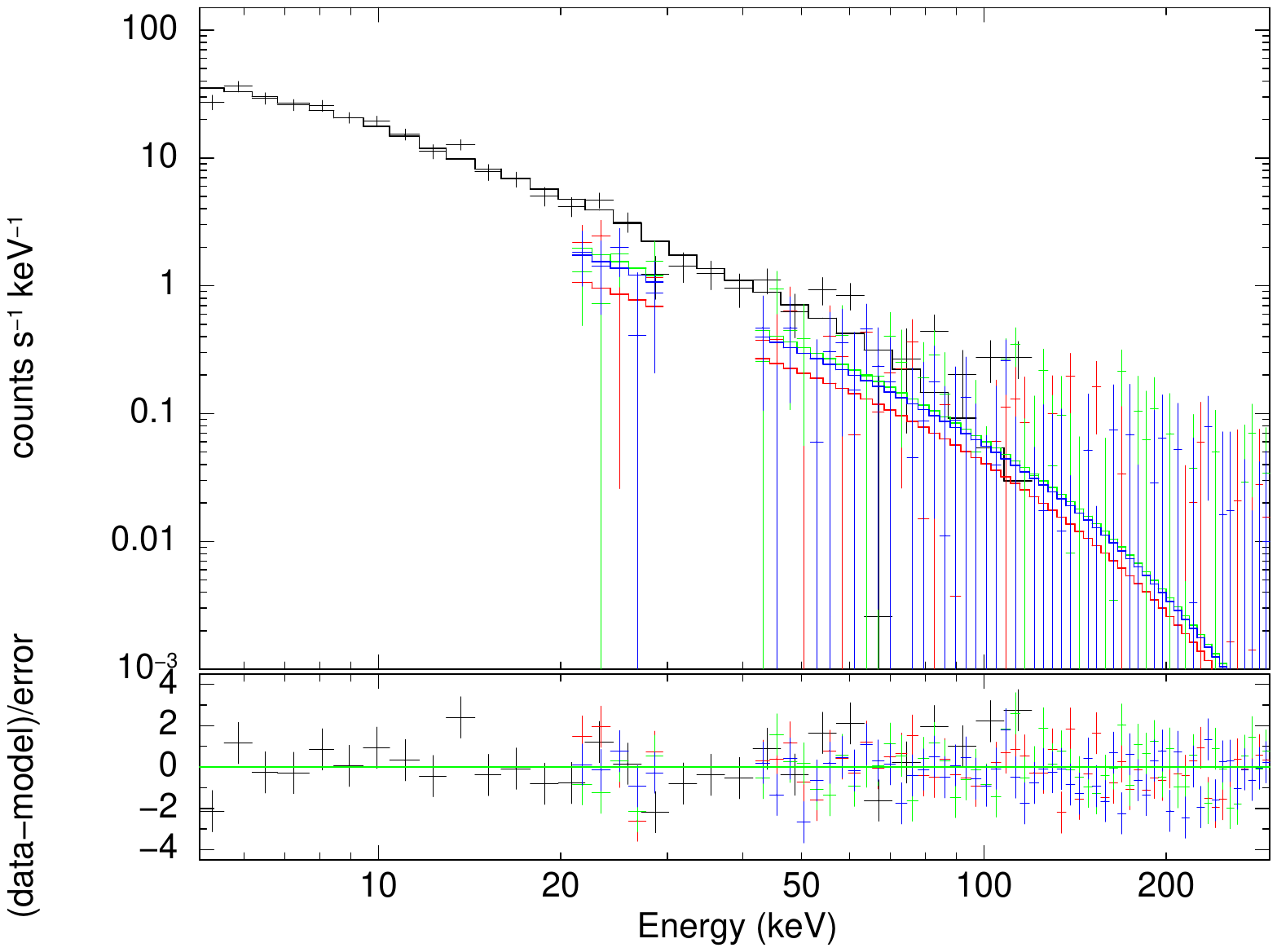}}  
    \vspace*{1.4ex}

    \caption{\textbf{GRB240821A, a fully characterised short GRB with extended emission: joint ECLAIRs+GRM spectral analysis.}
    Top: spectral analysis of the initial spike. The best fit model is a Band function with photon indices $\alpha_1=-0.68^{+0.07}_{-0.07}$ and $\alpha_2=-1.76^{+0.07}_{-0.08}$, and a break energy $E_\mathrm{b}=109^{+15}_{-19}\, \mathrm{keV}$. As the high-energy photon index is $\alpha_2>-2$, the peak energy cannot be measured  but is at higher energy than the detected break.
    Bottom: spectral analysis of the extended emission. The best fit model is a power-law with an exponential cutoff with a photon index $\alpha=-1.67^{+0.09}_{-0.09}$ and a peak energy $E_\mathrm{p}=25^{+12}_{-9}\, \mathrm{keV}$.
    Figures adapted from \citetalias{svom_paper_240821A_inprep}.
    In both time intervals, the spectral analysis is carried with the ECLAIRs pipeline (ECPI) for ECLAIRs data using a chi-2 statistics (black) and eclgrm-xband for the three GRDs of GRM using \textit{pgstat} (GRD1: red, GRD2: green, and GRD3: blue) \citep[see][]{Godet+etal+2026a,Goldwurm+etal+2026,Piron+etal+2026}.
    The analysis is carried out from 5 keV to 1 MeV for the initial spike, and to 300 keV for the extended emission.}
   \label{fig:ecl_grm_240821A_sp}
\end{figure}

\subsection{\textit{SVOM} Early View on Short GRBs}
\label{sect:short}

\begin{figure*}
   \centering
   \includegraphics[height=5cm]{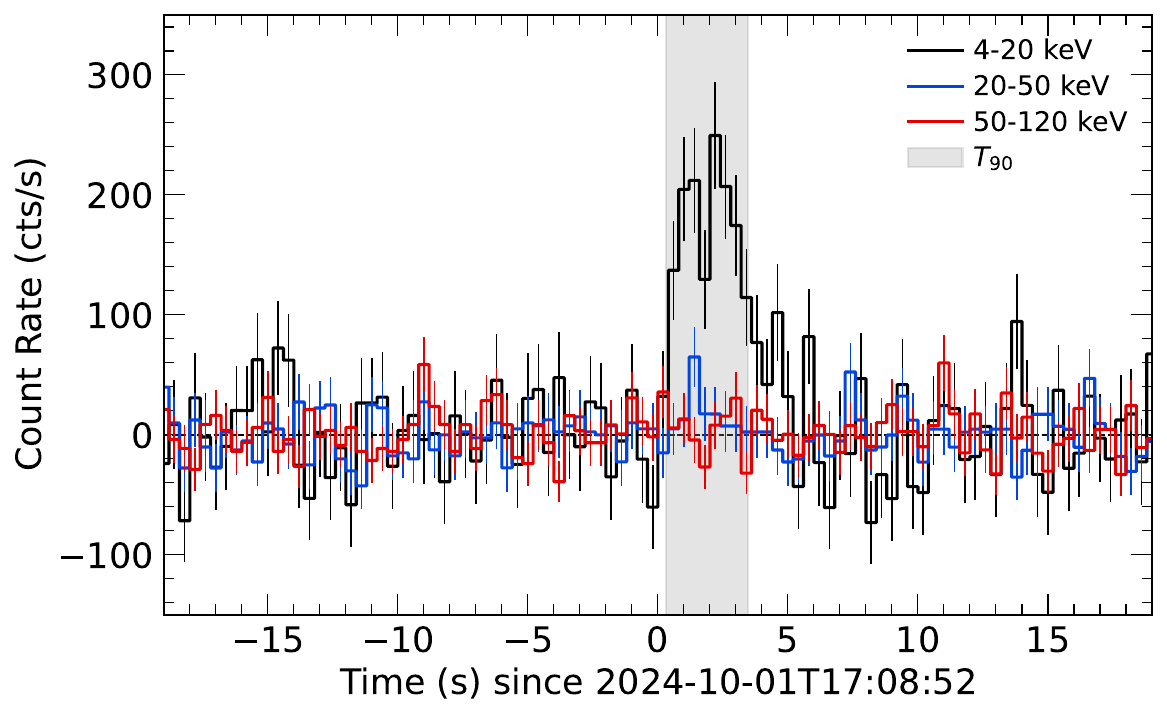}
   \hspace*{1cm}
   \includegraphics[height=5cm]{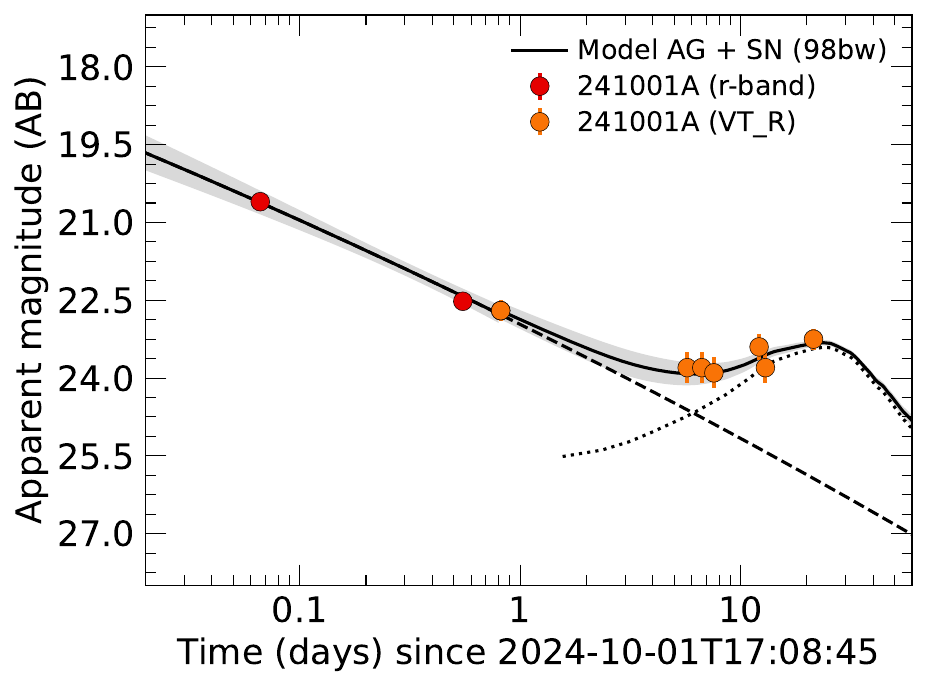}

   \caption{\textbf{GRB241001A, a fully characterised X-ray flash.}
   Left: ECLAIRs background subtracted lightcurve using only  pixels of the detector plane illuminated by the burst. 
   The burst is well detected in the 4-20 keV energy band (black) and not at higher energy.
   The shaded region corresponds to the interval of $T_{90}
= 3.14 \pm 0.18\, \mathrm{s}$ (4-20 keV).
   Right: visible afterglow lightcurve, observed by the Las Campanas Observatory (LCO) network, VLT/X-shooter
   and \textit{SVOM}/VT (see references in Table~\ref{tab:bestsample}). The lightcurve is modeled by the sum of an afterglow component and a supernova.
   The afterglow is fit by a powerlaw with a temporal decay index $\alpha =  0.81 \pm 0.06$.
   The supernova component corresponds to the type-Ic broad line supernova associated to GRB980425 \citep{galama:1998}, shifted to the redshift $z=0.573 $ of this GRB.
   Figures adapted from \citet{svom_paper_241001A}.}
   \label{fig:241001A}
\end{figure*}

It is believed that most short GRBs are type-I GRBs associated to binary neutron star mergers \citep{Zhang_2007}. Among them, a particular class of interest consists in short GRBs followed by an extended emission, as this additional episode of emission may bring new constraints on the post-merger evolution of the central source. 
A well studied example of such GRBs was detected by \textit{Swift} in 2006 \citep[GRB 060614,][]{Gehrels_2006} 
but the sample of well characterised events in this class remains small \citep[see e.g.][]{Kaneko_2015}. However, it is difficult to reach
a high level of confidence in a merger origin when only the prompt emission is available. It was
anticipated that \textit{SVOM} will increase the available sample as ECLAIRs is sensitive to the extended emission, which is usually softer than the initial short hard GRB \citep{svom_white_paper}. This is illustrated by the detection of GRB240821A. As seen in Figure~\ref{fig:ecl_grm_240821A_lc}, the lightcurve shows a first bright and variable episode lasting less than 2 s, well detected by ECLAIRs and GRM. This initial spike is followed by a softer emission for at least 50 seconds, 
still with possible variability.
As shown in Figure~\ref{fig:ecl_grm_240821A_sp}, this weaker emission is softer but remains non-thermal.
Thanks to an efficient follow-up, this GRB was fully characterised (see Table~\ref{tab:bestsample}). 
The ongoing analysis of the properties of the weak afterglow and of the host galaxy seems to point to a merger origin. This would make GRB240821A an ideal case for probing the post-merger physics by studying the properties of the extended emission. 
This will need to be confirmed in the forthcoming paper \citepalias{svom_paper_240821A_inprep}.

\subsection{\textit{SVOM} Early Exploration of the Softer GRBs}
\label{sect:soft}

Thanks to the low-energy threshold at 4 keV of ECLAIRs, it is expected that \textit{SVOM} will characterise the population of soft GRBs revealed by the Satellite per Astronomia X (Beppo-SAX, \citealt{Heise_2001}) and  the High Energy Transient Explorer 2 (\textit{HETE-2}, \citealt{Sakamoto_2005}) and currently also explored by \textit{EP} \citep[see e.g.][]{Gao_2025}. 
Depending on their spectral properties, these events are often classified as X-ray Rich GRBs (XRRs) and X-ray Flashes (XRFs) \citep{Barraud2003,Sakamoto2008}.
This characterisation is an important goal for \textit{SVOM}, as this population shows some diversity and their origin and precise connection to GRBs are highly debated. 
In particular, it is unclear whether the origin of the unusual properties of these GRBs is due to external  (high redshift, off-axis view) or intrinsic reasons (dirty fireball, shock breakout of a mildly relativistic jet, etc.). In the second case, the properties of the relativistic ejecta leading to a weak and soft emission remain to be understood. In some cases, it is even expected that the dominant mechanism for the prompt emission may be different, such as a shock breakout as in the example of XRF080109 \citep{Soderberg_2008}.

GRB241001A is a first example of a soft GRB detected by ECLAIRs that can be classified as an XRF. As shown in Figure~\ref{fig:241001A} (left), this weak event 
lasted several seconds with a variable lightcurve and was
detected only in the 4-20 keV energy band. 
Besides, the visible late emission
has a typical GRB afterglow evolution (see Figure~\ref{fig:241001A}, right). The bump detected by \textit{SVOM}/VT after 10 days is the signature of an associated type-Ic broad line supernova detected by JWST (see Table~\ref{tab:bestsample}). This clearly points to a progenitor similar to those of classical long GRBs.
As discussed in \citet{svom_paper_241001A}, this soft event would very likely not have been detected by \textit{Swift}/BAT. 
On the other hand the afterglow could in principle still be detected at least up to $z=1$ with the sensitivity of current telescopes.
It remains however significantly  fainter than the on-axis orphan afterglows detected by the Zwicky Transient Factory (ZTF) \citep{Ho_2022}, which may either indicate that GRB241001A belongs to a different population
than ZTF orphan afterglow or that the ZTF survey preferentially select bright afterglows.
Due to the weakness of the prompt emission, the spectral analysis does not allow to distinguish between a thermal or a non-thermal spectrum but shows that the peak energy is of the order of 7~keV and that the burst is under-luminous, with an isotropic equivalent energy $E_{\gamma,\mathrm{iso}}\simeq 7\times 10^{49}\, \mathrm{erg}$.
The analysis presented in \citet{svom_paper_241001A} points to a collapsar followed by an energetic relativistic ejection, but with low prompt emission efficiency.

\begin{figure*}
    \centering
    \begin{tabular}{cc}
        \begin{minipage}{0.5\linewidth}
        \vspace*{-0.3cm}
        \includegraphics[width=0.84\linewidth, angle=0]{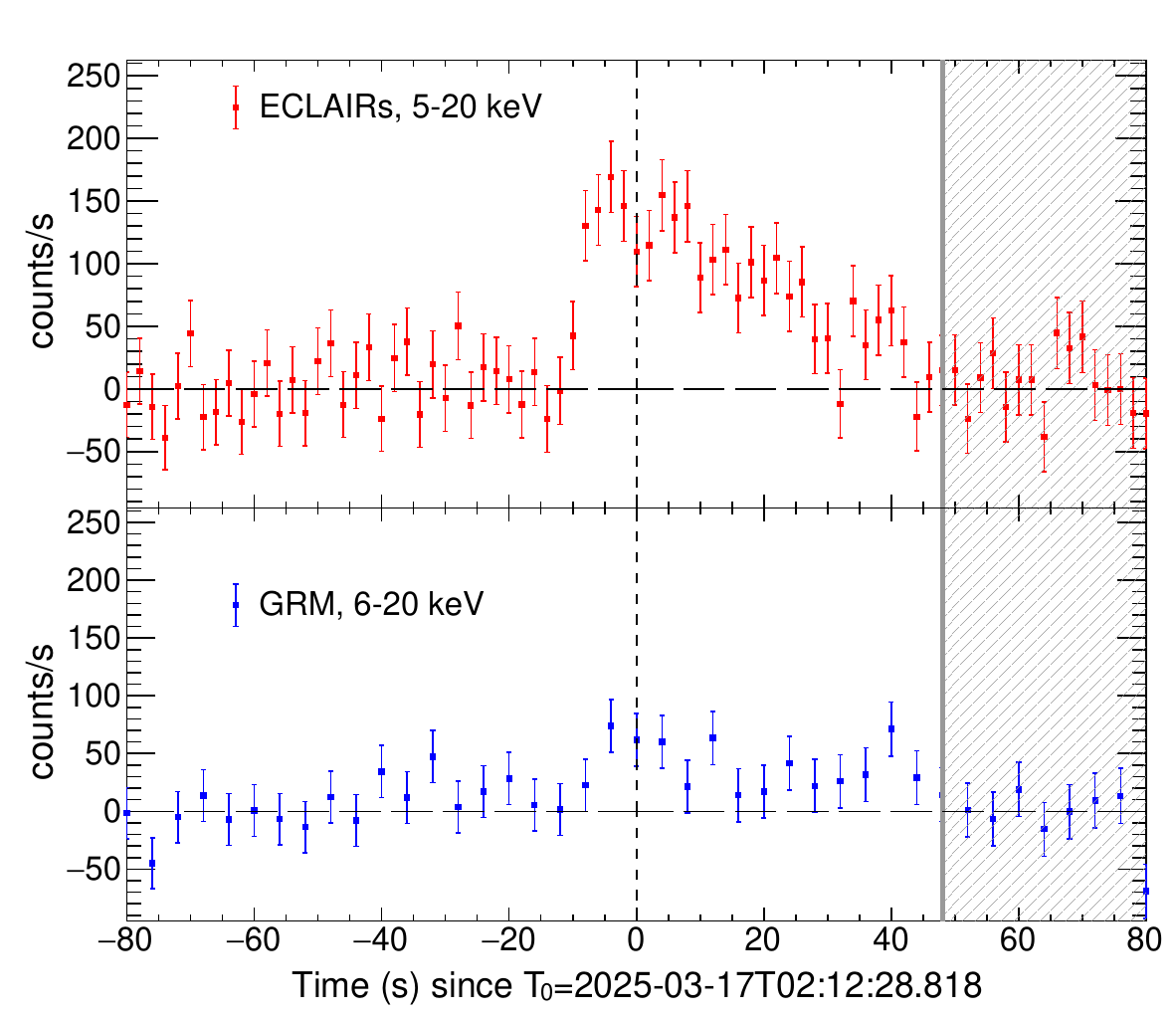}
        \end{minipage}
    &
        \begin{minipage}{0.5\linewidth}
        \includegraphics[width=0.83\linewidth, angle=0]{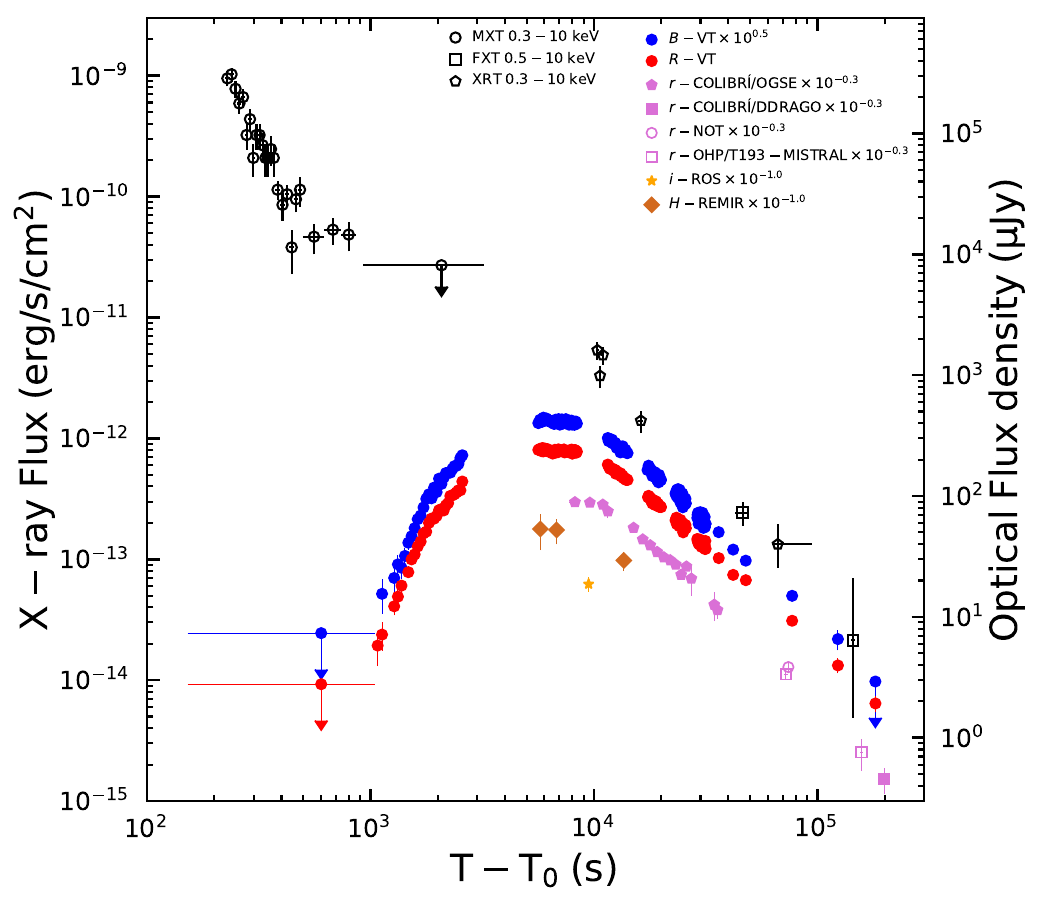}
        \end{minipage}\\
   \end{tabular}
   \caption{\textbf{GRB250317B, an X-ray Flash at large redshift with an unusual early afterglow.}
   Left: background-subtracted lightcurve in ECLAIRs (top) and GRM (bottom). 
   Right: afterglow lightcurve in X-rays (black symbols) and visible (colored symbols). In addition to the references in Table~\ref{tab:bestsample}, the observations by the NOT, the Multi-purpose InSTRument for Astronomy at Low-resolution (MISTRAL) at the T194 telescope of Observatoire de Haute Provence (OHP)
   and the Rapide Eye Mount telescope (REM, visible imager ROS and infrared imaging camera REMIR) 
   are taken from GCN~\citetalias{GCN_NOT_250317B,GCN_OHP_250317B,GCN_REM_250317B}. Figures adapted from \citetalias{svom_paper_250317B_inprep}.}
   \label{fig:250317B}
\end{figure*}

GRB250317B is a second example of an XRF detected by ECLAIRs. As seen in Figure~\ref{fig:250317B} (left), it is again a soft event detected mostly by ECLAIRs below 20 keV. As reported in Table~\ref{tab:bestsample} and Figure~\ref{fig:250317B} (right), this event benefited from an excellent multi-wavelength follow-up and has a the largest redshift measured for an XRF, $z=3.44$. 
The corresponding isotropic equivalent radiated energy is high, $E_\mathrm{\gamma,iso}\sim 1.8\times 10^{52}\, \mathrm{erg}$, despite a low peak energy in the source frame \citepalias{svom_paper_250317B_inprep}.
The early afterglow shows a peculiar behaviour, with a late rise of the visible afterglow, that peaks at $\sim 6000\, \mathrm{s}$ post-trigger, and then decays with a temporal decay index $\alpha\sim 1.7$. On the other hand, the X-ray afterglow starts with an early steep decay, followed by a plateau or shallow phase and then decaying with a temporal decay index close to that observed in the visible. As discussed in \citetalias{svom_paper_250317B_inprep}, the visible
and late X-ray afterglow lightcurves 
may indicate that the external medium
is highly structured, which would then put new constraints on the mass loss history of the progenitor star during the thousand years before the explosion.
Hence, this event is quite different from the previous example of XRF, GRB241001A, 
which illustrates
the diversity of the soft GRB population.
SVOM observations of these two first  examples are promising in terms of its ability to precisely characterize such very soft GRBs.

\subsection{\textit{SVOM} and the Population of Classical Long GRBs}
\label{sect:long}

As shown in Table~\ref{tab:bestsample}, \textit{SVOM}  detects and characterises many classical long GRBs, a good fraction of them in common with other satellites. 
Two examples have already been discussed in Section~\ref{sec:GRM-only_GRBs}.
In particular, GRB250129A in Figure~\ref{fig:vt_250129A_241105A} (left) illustrates how \textit{SVOM} allows an excellent characterisation of the visible aferglow, often revealing a surprising variability that is challenging for the standard external forward shock model. 

\begin{figure}
    \centering
    \includegraphics[width=0.9\linewidth]{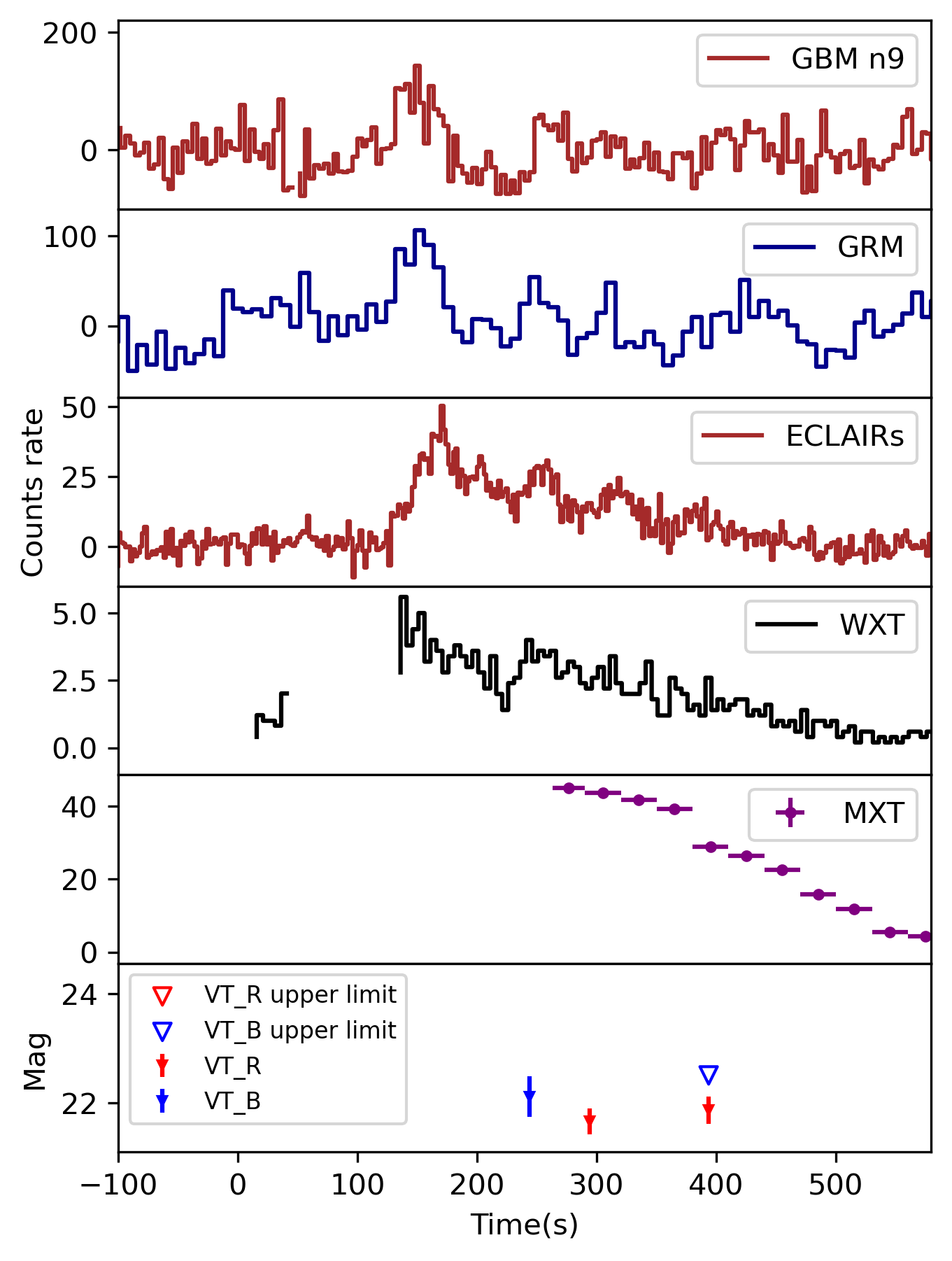}
    \caption{\textbf{GRB241217A, a very long GRB with an excellent multi-wavelength coverage from the end of the prompt emission to the afterglow phase: (1) prompt emission.} Background-subtracted lightcurve in \textit{Fermi}/GBM, \textit{SVOM}/GRM, \textit{SVOM}/ECLAIRs and \textit{EP}/WXT. The two last panels show the \textit{SVOM}/MXT and \textit{SVOM}/VT lightcurves obtained after the slew of the satellite. See references in Table~\ref{tab:bestsample}. Figure adapted from \citetalias{svom_paper_241217A_inprep}.}
    \label{fig:241217A_prompt}
\end{figure}

\begin{figure}
    \centering
    \includegraphics[width=0.9\linewidth]{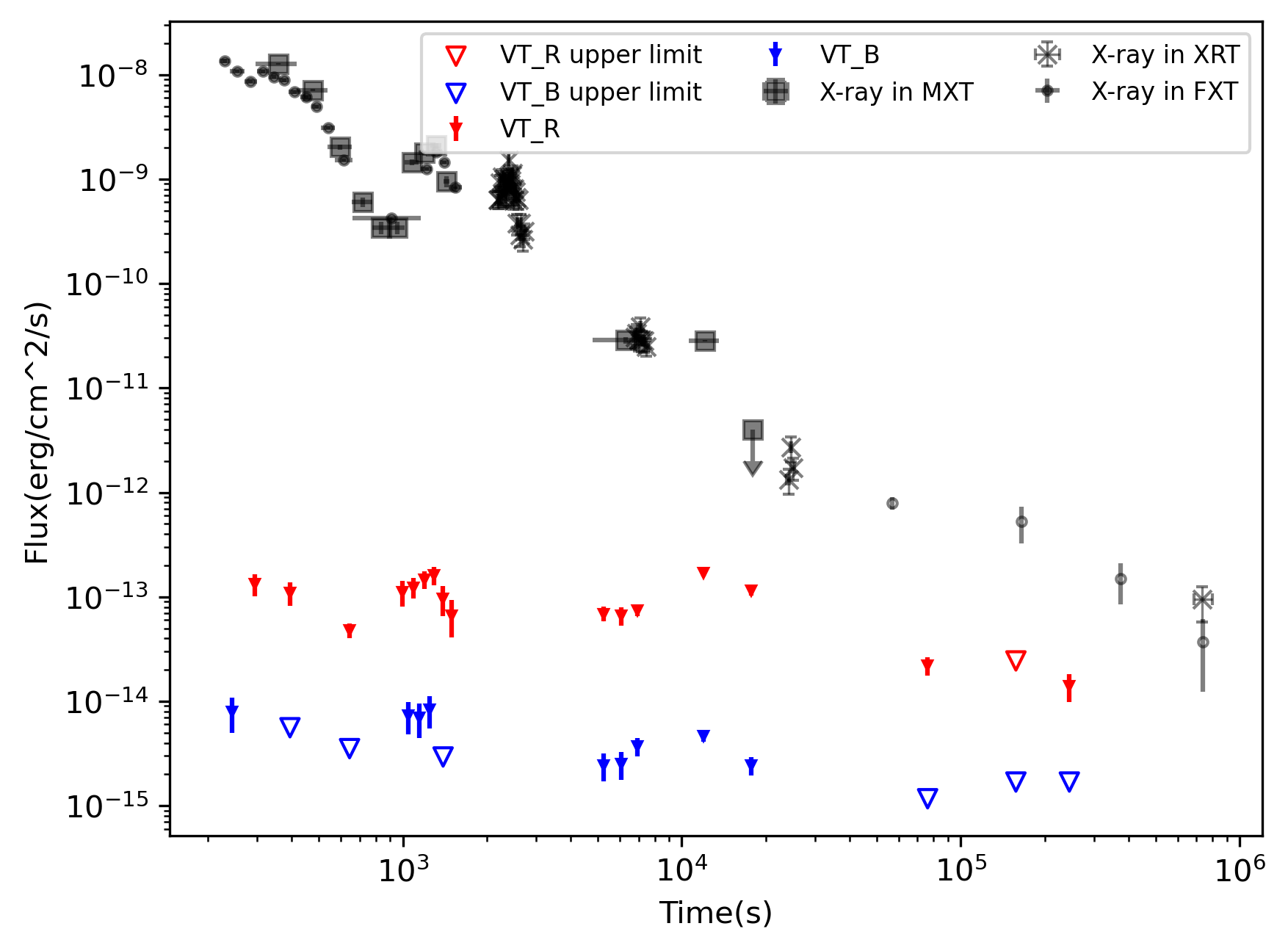}
    \caption{\textbf{GRB241217A, a very long GRB with an excellent multi-wavelength coverage from the end of the prompt emission to the afterglow phase: (2) afterglow.} Lightcurve in X-rays (black symbols) and visible (red and blue symbols), see references in Table~\ref{tab:bestsample}. Figure adapted from \citetalias{svom_paper_241217A_inprep}.}
    \label{fig:241217A_afterglow}
\end{figure}

The diversity and variability of the afterglows of \textit{SVOM} long GRBs is also illustrated with the especially interesting case of 
GRB241217A. This burst was not only detected  by \textit{SVOM}/ECLAIRs and GRM, but also by \textit{Fermi}/GBM and \textit{EP}/WXT (see Table~\ref{tab:bestsample}). 
As shown in Figure~\ref{fig:241217A_prompt}, the end of its very long prompt emission was detected in the X-ray and visible bands with MXT and VT,
thanks to the early automatic slew of the \textit{SVOM} satellite, before the transition to the afterglow. 
The latter shows a very peculiar behavior, especially in the visible range, as shown by the VT lightcurve in Figure~\ref{fig:241217A_afterglow}. It starts with a very shallow decay phase up to at least $\sim 10^4\, \mathrm{s}$ before probably recovering a normal decay, however difficult to characterise due to the very low flux at that stage. The early plateau-like phase shows 
a flare at $\sim 10^3\, \mathrm{s}$ and possibly a second one at $\sim 10^4\, \mathrm{s}$,
very probably also seen in X-rays. As discussed in \citetalias{svom_paper_241217A_inprep}, this very unusual behaviour is particularly difficult to interpret in the standard scenario. One possible explanation is that the first phase is dominated by late central engine activity, observed for an exceptionally long duration due to a low-density external environment delaying the normal afterglow.

In addition to such afterglow observations, 
the combination of ECLAIRs and GRM will also enable detailed spectral analysis of the prompt emission of long GRBs \citep{Bernardini_2017}. In particular, the low-energy threshold at 4 keV of ECLAIRs allows to better constrain the shape of the soft gamma-ray spectrum, 
(see the example of the detection of a low-energy break in GRB250506A, Figure~13 in \citealt{Godet+etal+2026a}). This is especially important as different scenarios for the prompt emission (emission of a dissipative photosphere or synchrotron radiation from accelerated electrons above the photosphere) have very different predictions in this energy range.

Overall, the sample of long GRBs detected by \textit{SVOM} will be particularly useful in advancing our understanding of their different emission phases, thanks to the precise characterisation of their prompt emission and the efficient and systematic multi-wavelength afterglow follow-up and redshift measurement (see Section~\ref{sec:characterization} and Table~\ref{tab:followup_efficiency}).

 \begin{figure}
    \centering
    \includegraphics[width=0.46\linewidth]{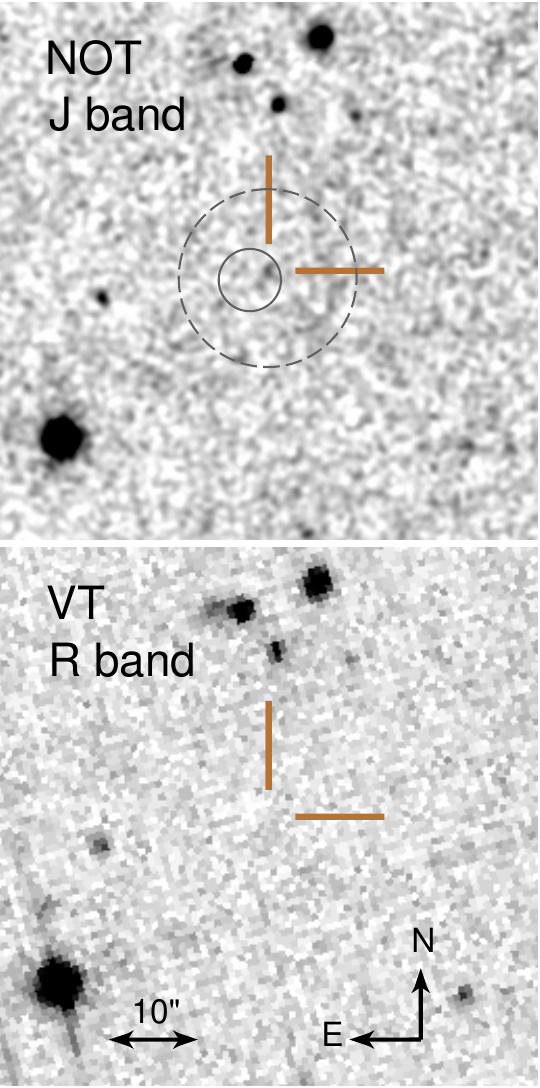}
    \caption{\textbf{GRB250314A, a classical long GRB at high redshift ($z\simeq 7.3$): afterglow detection.}
    Bottom: VT image in R band, obtained by stacking 23 images from the first \textit{SVOM} orbit following the trigger. No counterpart is identified, leading to an early deep upper limit in the red filter. 
    Top: NOT image in J band 12.1~h after the trigger. The NIR counterpart is identified by the two copper-colored lines, well within the X-ray error circle of \textit{EP}/FXT (dashed line) and \textit{Swift}/XRT (solid line). Both images have an approximate size of $1'\times 1'$. The NOT image was smoothed. The bottom figure is adapted from \citet{svom_paper_250314A}.}
   \label{fig:250314A_images}
\end{figure}

\subsection{\textit{SVOM} GRBs as Probes of the Distant Universe}
\label{sect:cosmo}

An important objective of the \textit{SVOM} core program is the detection of high-redshift GRBs to probe the distant Universe \citep{svom_paper_250314A}. The low-energy threshold of ECLAIRs  is expected to favor the detection of such events \citep[see e.g.][]{Palmerio_2021}, and the sensitivity in two channels of VT, including the VT\_R channel extending to 1000~nm, allows the early identification of high-redshift candidates \citep{Wang_2020}.

The detection of GRB250314A 
at $z\simeq 7.3$ is a first major success in achieving this scientific goal \citep{svom_paper_250314A}. 
This is the third highest spectroscopic redshift ever obtained for a GRB.
This brings the sample size of GRBs beyond $z=7$ to five, with the previous ones dating from 2009 to 2012 (GRB120923A at $z\simeq 7.8$, \citealt{Tanvir_2018}; 
GRB100905A at $z=7.88$, \citealt{Im_2012};
GRB090423A at $z=8.23$, \citealt{Tanvir_2009,Salvaterra_2009};
and GRB090429B at $z\simeq 9.4$, \citealt{Cucchiara_2011}).
As expected at such a high redshift, the afterglow was not detected by VT, providing a deep upper limit, but it was detected in the NIR by the NOT telescope (see Figure~\ref{fig:250314A_images}).

\begin{figure}
    \centering
    \includegraphics*[width=0.9\linewidth,viewport=0.1cm 0cm 14.9cm 11cm]{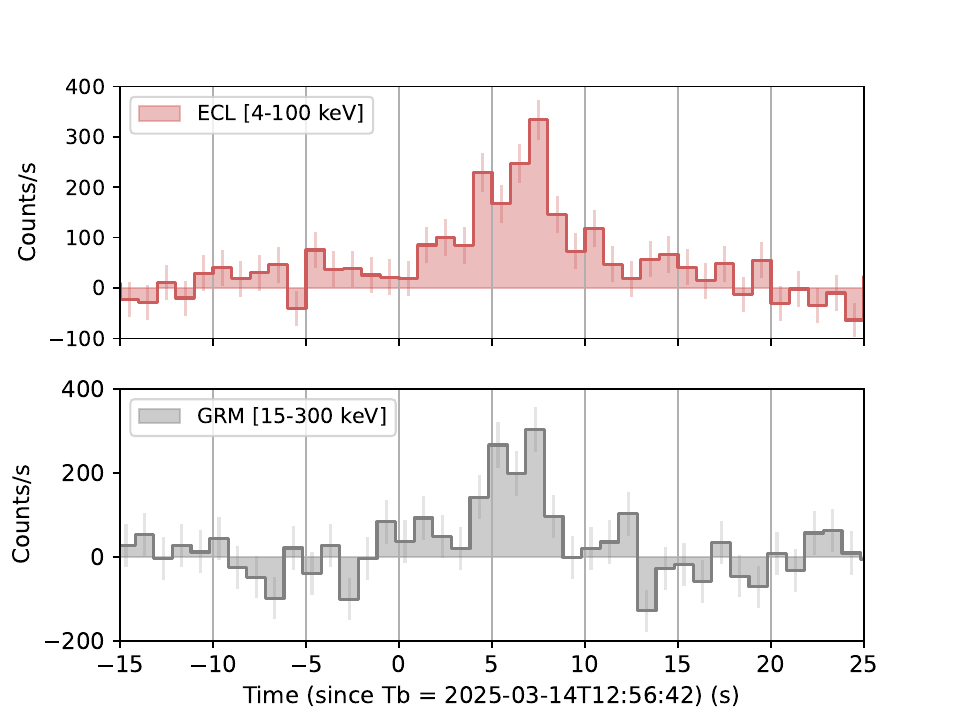}  
    \caption{\textbf{GRB250314A, a classical long GRB at high redshift ($z\simeq 7.3$): prompt emission.}
    Background-subtracted lightcurves in ECLAIRs in the 4–100 keV energy band (using only pixels of the detector plane illuminated by the burst, top panel) and in GRM in the 15–300 keV energy band (bottom panel) using a time bin of 1~s. Figure adapted from \citet{svom_paper_250314A}.}
    \label{fig:250314A_prompt}
\end{figure}

The prompt ECLAIRs and GRM lightcurves are shown in Figure~\ref{fig:250314A_prompt}.
Despite a short observed duration in the source rest frame, $T_{90}/(1+z)=1.3^{+0.4}_{-0.2}\, \mathrm{s}$, which can be interpreted as a "tip of the iceberg" effect expected at such redshifts \citep{Lu2014,Llamas_2024}, all other properties, including the location of this GRB in the Amati diagram, points towards a type-II GRB associated to the collapse of a massive star \citep{svom_paper_250314A}. This is directly confirmed by JWST observations 110 days after the trigger (13 days rest-frame), which reveal an associated supernova compatible with a type-Ic broad line supernova
similar to those associated to long GRBs in the local Universe \citep{JWST_GRB250314A}.
This proves the capacity to observe with \textit{SVOM} the explosion of an individual massive star during the epoch of reionisation.  

To fully exploit the scientific potential of such high-redshift GRBs, which remain relatively rare, the follow-up strategy still needs to be optimized. Indeed, the VLT/X-shooter
spectra were  obtained $\sim 17$~h after the trigger in the case of GRB250314A, and 
only allow for a redshift measurement, based on the Lyman-$\alpha$ break. 
\citet{svom_paper_250314A} discuss in details how this delay could  be significantly reduced for future similar events, to perform spectroscopy when the source is still
bright and to detect metal lines.
This implies in particular to make the deep upper limits obtained by VT available much earlier, 
which can be achieved with 
more X-band stations to get all acquired images 
faster. This also implies to improve NIR follow-up capabilities,
which should be achieved in 2026 with the start of the SOXS spectrograph (Son of X-Shooter, \citealt{Schipani_2018}) and the commissioning of the CAGIRE camera at COLIBRI  \citep{Fortin_2024}.

\section{Conclusion}
\label{sect:conclusion}

The first 9.3 months of the \textit{SVOM} mission have already seen the detection of 86 GRBs.
This article illustrates how this sample in construction already explores a wide diversity of GRBs.
There is already a clear impact of 
(i) the low-energy threshold of ECLAIRs at 4~keV and its capacity to trigger on long timescales. In particular, this gives access to the softer end of the GRB population. In combination with the broad spectral range of the GRM, it also allows for a full characterisation of the prompt spectrum; 
(ii) the optimization of the follow-up sequence, especially with an anti-solar pointing, the excellent sensitivity of VT, the complementary observations by the C-GFT and COLIBRI, and strong partnerships with many other instruments. The synergy with \textit{Swift}/XRT and \textit{EP}/FXT to complement MXT X-ray afterglow observations is crucial. The  excellent efficiency of the rate of visible/NIR afterglow detections and redshift measurements during the first 9.3 months is promising for the capacity of \textit{SVOM} to build a sample of fully characterised GRBs. 
This is illustrated in this article with several examples, including a short GRB with extended emission, two XRFs, and several classical long GRBs including a high-redshift one. These first detected GRBs are discussed in more details in forthcoming papers.

\begin{acknowledgements}
The Space-based multi-band astronomical Variable Objects Monitor (\textit{SVOM}) is a joint Chinese-French mission led by the Chinese National Space Administration (CNSA), the French Space Agency (CNES), and the Chinese Academy of Sciences (CAS). We gratefully acknowledge the unwavering support of NSSC, IAMCAS, XIOPM, NAOC, IHEP, CNES, CEA, and CNRS.

SDV, BS, AS and SB acknowledge the support of the French Agence Nationale de la Recherche (ANR), under grant ANR-23-CE31-0011 (project PEGaSUS).

AS acknowledges financial support from the Centre national d’études spatiales (CNES), France (ROR: \url{https://ror.org/04h1h0y33}) within the framework of the SVOM mission.

This publication made use of data from the Legacy Surveys.
The Legacy Surveys consist of three individual and complementary projects: the Dark Energy Camera Legacy Survey (DECaLS; Proposal ID \#2014B-0404; PIs: David Schlegel and Arjun Dey), the Beijing-Arizona Sky Survey (BASS; NOAO Prop. ID \#2015A-0801; PIs: Zhou Xu and Xiaohui Fan), and the Mayall z-band Legacy Survey (MzLS; Prop. ID \#2016A-0453; PI: Arjun Dey). DECaLS, BASS and MzLS together include data obtained, respectively, at the Blanco telescope, Cerro Tololo Inter-American Observatory, NSF’s NOIRLab; the Bok telescope, Steward Observatory, University of Arizona; and the Mayall telescope, Kitt Peak National Observatory, NOIRLab. Pipeline processing and analyses of the data were supported by NOIRLab and the Lawrence Berkeley National Laboratory (LBNL). The Legacy Surveys project is honored to be permitted to conduct astronomical research on Iolkam Du’ag (Kitt Peak), a mountain with particular significance to the Tohono O’odham Nation.

NOIRLab is operated by the Association of Universities for Research in Astronomy (AURA) under a cooperative agreement with the National Science Foundation. LBNL is managed by the Regents of the University of California under contract to the U.S. Department of Energy.

This project used data obtained with the Dark Energy Camera (DECam), which was constructed by the Dark Energy Survey (DES) collaboration. Funding for the DES Projects has been provided by the U.S. Department of Energy, the U.S. National Science Foundation, the Ministry of Science and Education of Spain, the Science and Technology Facilities Council of the United Kingdom, the Higher Education Funding Council for England, the National Center for Supercomputing Applications at the University of Illinois at Urbana-Champaign, the Kavli Institute of Cosmological Physics at the University of Chicago, Center for Cosmology and Astro-Particle Physics at the Ohio State University, the Mitchell Institute for Fundamental Physics and Astronomy at Texas A\&M University, Financiadora de Estudos e Projetos, Fundacao Carlos Chagas Filho de Amparo, Financiadora de Estudos e Projetos, Fundacao Carlos Chagas Filho de Amparo a Pesquisa do Estado do Rio de Janeiro, Conselho Nacional de Desenvolvimento Cientifico e Tecnologico and the Ministerio da Ciencia, Tecnologia e Inovacao, the Deutsche Forschungsgemeinschaft and the Collaborating Institutions in the Dark Energy Survey. The Collaborating Institutions are Argonne National Laboratory, the University of California at Santa Cruz, the University of Cambridge, Centro de Investigaciones Energeticas, Medioambientales y Tecnologicas-Madrid, the University of Chicago, University College London, the DES-Brazil Consortium, the University of Edinburgh, the Eidgenossische Technische Hochschule (ETH) Zurich, Fermi National Accelerator Laboratory, the University of Illinois at Urbana-Champaign, the Institut de Ciencies de l’Espai (IEEC/CSIC), the Institut de Fisica d’Altes Energies, Lawrence Berkeley National Laboratory, the Ludwig Maximilians Universitat Munchen and the associated Excellence Cluster Universe, the University of Michigan, NSF’s NOIRLab, the University of Nottingham, the Ohio State University, the University of Pennsylvania, the University of Portsmouth, SLAC National Accelerator Laboratory, Stanford University, the University of Sussex, and Texas A\&M University.

BASS is a key project of the Telescope Access Program (TAP), which has been funded by the National Astronomical Observatories of China, the Chinese Academy of Sciences (the Strategic Priority Research Program “The Emergence of Cosmological Structures” Grant \# XDB09000000), and the Special Fund for Astronomy from the Ministry of Finance. The BASS is also supported by the External Cooperation Program of Chinese Academy of Sciences (Grant \# 114A11KYSB20160057), and Chinese National Natural Science Foundation (Grant \# 12120101003, \# 11433005).

The Legacy Survey team makes use of data products from the Near-Earth Object Wide-field Infrared Survey Explorer (NEOWISE), which is a project of the Jet Propulsion Laboratory/California Institute of Technology. NEOWISE is funded by the National Aeronautics and Space Administration.

The Legacy Surveys imaging of the DESI footprint is supported by the Director, Office of Science, Office of High Energy Physics of the U.S. Department of Energy under Contract No. DE-AC02-05CH1123, by the National Energy Research Scientific Computing Center, a DOE Office of Science User Facility under the same contract; and by the U.S. National Science Foundation, Division of Astronomical Sciences under Contract No. AST-0950945 to NOAO.

The Pan-STARRS1 Surveys (PS1) and the PS1 public science archive have been made possible through contributions by the Institute for Astronomy, the University of Hawaii, the Pan-STARRS Project Office, the Max-Planck Society and its participating institutes, the Max Planck Institute for Astronomy, Heidelberg and the Max Planck Institute for Extraterrestrial Physics, Garching, The Johns Hopkins University, Durham University, the University of Edinburgh, the Queen's University Belfast, the Harvard-Smithsonian Center for Astrophysics, the Las Cumbres Observatory Global Telescope Network Incorporated, the National Central University of Taiwan, the Space Telescope Science Institute, the National Aeronautics and Space Administration under Grant No. NNX08AR22G issued through the Planetary Science Division of the NASA Science Mission Directorate, the National Science Foundation Grant No. AST–1238877, the University of Maryland, Eotvos Lorand University (ELTE), the Los Alamos National Laboratory, and the Gordon and Betty Moore Foundation.
\end{acknowledgements}

\bibliographystyle{raa} 
\bibliography{bibtex.bib}  
\end{document}